\documentclass[12pt]{article}
\usepackage{graphicx}
\begin{document} 

\centerline{\it For Physics of Life Reviews}

\bigskip
\centerline{\bf
Microscopic and Macroscopic Simulation}

\centerline {\bf of Competition between Languages}

\bigskip

\centerline{Dietrich Stauffer and Christian Schulze}

\centerline{Institute for Theoretical Physics, Cologne University, D-50923 K\"oln, Euroland}

\bigskip
\centerline{e-mail: stauffer@thp.uni-koeln.de}

\bigskip
Abstract: 
The similarity of the evolution of human languages (or alphabets, bird songs, 
...) to biological evolution of species is utilized to study with up to $10^9$
people the rise and fall of languages either by macroscopic differential 
equations similar to biological Lotka-Volterra equation, or by microscopic 
Monte Carlo simulations of bit-strings incorporating the birth, maturity, and 
death of every individual. For our bit-string model, depending on parameters 
either one language comprises
the majority of speakers (dominance), or the population splits into many 
languages having in order of magnitude the same number of speakers 
% This sentence new, replaces old one
(fragmentation); in the latter case the size distribution is log-normal, with 
upward deviations for small sizes, just as in reality for human languages.
On a lattice two different dominating languages can coexist in neighbouring
regions, without being favoured or disfavoured by different status. 
We deal with modifications and competition for existing languages, 
not with the evolution or learning of one language.

\bigskip


\section{Introduction}

In biological just as in physical computer simulations, one may study phenomena
microsopically or macroscopically. Microscopic physics simulations use 
Monte Carlo or Molecular Dynamics methods since half a century e.g. to
study single atoms. Macroscopic simulations average over many atoms and study
their average properties e.g. by differential equations or other mean field
approximations; hydrodynamics or reaction equations for the concentrations
of chemical compounds are famous examples. Similarly, in biology we may either
look at the birth, the ageing, and the death of every single individual in 
a microscopic study. Or we may write down and solve differential equations 
for the total number of individuals in a macroscopic description, like in the 
Lotka-Volterra equations for prey and predator. P{\c e}kalski \cite{pekalski}
reviewed how these prey-predator relations can be studied also microscopically,
and Ref. \cite{kunwar} summarizes many recent microscopic and macroscopic 
simulations of ecosystems. In the socio-economic sciences, microscopic studies
may be called ``agent-based''.

One important aspect of a microscopic study is the finite lifetime of many 
finite populations, which live infinitely long in an averaged macroscopic
approach. Imagine we start with $N$ individuals, and at each time step 
randomly the number of individuals is either increased by one or decreased
by one. On average, therefore, the number stays constant. But microscopically,
it exhibits a random walk, will become zero after some time,
and once it becomes zero, the population is extinct forever \cite{redner}. 
Only for an infinitely large population also the
typical extinction time is infinitely large.

This does not mean, however, that macroscopic (mean-field) descriptions 
become correct and agree with microscopic ones for sufficiently large
populations. The one-dimensional Ising chain with nearest-neighbour interactions
at positive temperatures has no 
phase transition and no spontaneous magnetization, even though in a mean 
field approximation it has both. Closer to biology, Shnerb et al
\cite{shnerb} simulated a two-species model on a lattice where both 
species survive even though one may vanish in an approximation by differential
equations. Thus, in general, we mistrust averaged descriptions through
differential equations; if everybody can interact with everybody independent 
of spatial distance, then a macroscopic description may become exact for 
infinite populations, at least if a noise term is added. 

The present review deals with languages and thus one may ask what it has to do
with biology and physics. Once the simulation methods are explained, the answer
will be obvious: The methods are the same even if the aims are different.
Instead of prey and predator in biology, we study two or more languages here, 
and they may be represented by a bit-string just like many models in biology
\cite {eigen,farmer,penna}. After a discussion of the problems to be solved
we first look at recent macroscopic simulations of others and then at 
microscopic Monte Carlo simulations.

\section{Problems}

The very old idea to describe language evolution as similar to the evolution
of biological species was quantified recently by Sutherland \cite{sutherland}, 
who also cites more literature on empirical facts on languages. A collection
of recent reviews on languages is published in \cite{science}. 

What is a language? Is it a dialect with a navy and an army behind it, as
remarked by linguist Max Weinreich (according to our referee)?
Spanish and Portuguese are regarded as different languages but both speakers
may converse better in ``Portu\~nol'' than speakers from the northern-most
islands of Germany and those from the southern-most mountains of Germany, if 
they both use their dialect of German. Many nations have more than one official 
language, and many different nations may speak one and the same language 
though in different dialects. Similar problems arise in the biological taxonomy;
the definition of species through the ability to have viable offspring together
fails for asexual bacteria, where one may have to use an arbitrary percentage 
of DNA agreement to group bacteria into species. Nevertheless, there seems to 
be a widespread agreement that today humans speak several thousand languages,
of which many are in danger of extinction \cite{sutherland}.

The size $S_i$ of a language $L_i$ is the number of people speaking it, and the
size histogram $P(S)$, giving the number of languages of a given size,
is roughly log-normal, with enhanced numbers for small sizes \cite{sutherland}.
  
For the purposes of computer simulation, a language may also be a bird song,
a sign language, or an alphabet, that means we simulate ways in which
living beings communicate with each other. 

While the evolution of biological species may have taken many millions of 
years and can only partially be studied through the palaeontological record,
that of languages at least in some cases is much better documented: Within the
last two millennia, ancient Latin split into Portuguese, Spanish, French, 
Italian, Rumanian, ...; and again the French spoken in the Provence
differs from that in Paris or Haiti. Differently from biology, one language may
take over elements of another language for objects originating with the people
speaking that other language. And the English language is a merger of the
French spoken by the Norman conquerors of 1066 and the German spoken by the 
earlier Anglo-Saxon immigrants; such a merger does not happen between two 
biological species. 

Generally, simulation of language competition is still in its infancy, and we do
not deal here with computer-assisted analysis of existing real languages, 
the learning of languages by children, or the emergence of a human 
proto-language out of simpler sounds. Instead, we restrict ourselves to the few 
recent papers we know on simulating the evolution of adult populations speaking 
existing languages; simulations of the 1990's were reviewed by Livingstone
\cite{cang}. 

Models are supposed to describe the essential aspects of reality in a way as
% Whole paragraph added
simple as possible. In the present case this could mean that one looks only
at two languages, even though we know than many more are spoken by humans today.
Four centuries ago, Kepler approximated the planet Earth by a point when he
claimed that it runs around the sun on an ellipse. Clearly, he knew that the
Earth is not a point; but for the purpose of planetary motion, not for 
geography, this approximation is good. It lead to Newton's laws of motion and
the classical mechanics, but it took three centuries before the first airplanes
flew. Similarly, the models in this review are mainly of methodological value;
we should be happy if they give roughly the desired size distribution of human
languages and should not yet expect them to clarify, if the Indo-European 
languages \cite{comrie} originated $10^4$ years ago in Turkey's Konya plain, or 
later in Russia's Kurgan. 

\section{Macroscopic Simulations}

\subsection{Two Languages}

\subsubsection{Differential equations}

Since about two centuries, the growth of biological populations is described
by differential equations, where 
$$ dN/dt \propto N$$
leads to exponential growth (Malthus), while 
$$ dN/dt \propto N(1-N/K)$$
leads to saturation (Verhulst). Here $N(t)$ is the time-dependent population
size which for the Verhulst case approaches for long times the carrying 
capacity $K$; the factor of proportionality is supposed to be positive and 
related to the difference between birth rate and intrinsic death rate per 
person. With the abbreviation $x = N/K$ and suitable units for the time, the 
Verhulst equation gets the simple form
$$dx/dt = x(1-x) \quad .$$

Two species are described since many decades by the Lotka-Volterra equations
$$dx/dt = x-xy, \quad dy/dt = xy-{\rm const}\, y$$
for a prey population $x$ and a predator population $y$, with both population 
sizes and the time in suitable units to avoid more free parameters.

Theoretical biology often though unrealistically assumes the population of
a species to be constant \cite{eigen}, and then may use similar differential
equations to look for the fractions $x,\; y, \; ...$ of the population having
certain properties. This method is close to the differential equations to be 
used in this section for language evolution. 

In all these differential equations, the fate of the individual is ignored and
only their total number is studied, which is then approximated by a real number
instead of a natural number. (We interpret this as meaning that homo 
neanderthalis counts only a fraction of homo sapiens, but D\"usseldorf people 
may disagree.)

\subsubsection{Abrams-Strogatz Model}

The simplest computer simulation for language evaluation known to us was made
by Abrams and Strogatz \cite{strogatz} who used
$$ dx/dt = (1-x)x^as - x(1-x)^a(1-s)$$
for the fraction $x$ of people in a constant population, where everybody speaks
one of two languages X and Y. (We absorbed their free coefficient $c$ in our 
time unit.) The free parameter $s$ with $O < s < 1$ gives the relative status 
(=usefulness, prestige, ...) of language X, and $a$ is a positive free 
exponent. The relative status of the other language Y is $1-s$, and its fraction
of the population is $y=1-x$. The above equation
thus describes how people switch from one language to the other language. Thus
speakers from language Y switch over to language X with a rate per Y speaker
proportional to $x^as$, while the opposite switch from X to Y occurs with a rate
per X speaker of $(1-x)^a(1-s)$, i.e. in a complementary way. The fraction
of X speakers changes by the difference between influx and outflux. The factors
$x^a$ and $y^a$ take into account that people prefer to switch to a widely 
spoken language from a rare language. In biology,
the parameter $s$ would be called a relative fitness.

This model predicts that one language will die out and the other will be spoken
by everybody. More quantitatively, the number of people in Wales speaking Welsh,
and in Scotland speaking Gaelic (both competing with English), and in Peru 
speaking Quechua instead of Spanish, could be described over the 20th 
century by the above equation with same $a = 1.31 \pm 0.25$ and only slightly
different $s$ between 0.26 and 0.46. (For Guatemala see \cite{hawkins}.) 
With an active feedback term for $s$, also
bilingual communities like Canada (French minority and English majority) 
can be simulated as stable.

Fig.1 shows how in this model a disadvantaged language may vanish. Part a
% whole paragraph added
gives the symmetric case when we start with two equally strong languages, i.e.
$x(t=0)=1/2$. For small $s$ the fraction $x$ decays rapidly, for $s=1/2$ the
two languages are equally strong and nothing changes: $x(t) = 1/2$. For
larger $s$, $x$ approaches unity in a way symmetric to its approach towards zero.
The asymmetric case $x(t=0) < 1/2$ makes $x$ decaying towards zero even for 
$x = 0.5$ and 0.6; only for $s = 0.7$ and above does $x$ approach unity. 
In summary, with one exception always one of the two languages wins, and this
is the language which is favoured by a larger initial population or a more 
favourable status $s$.

\subsubsection{Patriarca-Lepp\"anen Geography}

Patriarca and Lepp\"anen \cite{finland}
put the Abrams-Strogatz model onto a square lattice such
that speakers can move from one site to a neighbour site. This diffusion process
alone may not have changed the results since as many speakers can move, on
average, from site A to site B as in the opposite direction from B to A. But
in addition the authors assumed that in half of the lattice, language X has
the higher status $s$, and in the other half the other language Y has the 
higher status. With this geographical difference of the status, they obtained
that each language dominated in the region where it was favoured, and no 
language vanished in the whole system. A detailed investigation of the interface
region, where the two languages mix in space, would be interesting, at least
for physicists worried about surface roughening etc. 

\subsection{Many Languages: Differential Equations}

The review of Nowak et al \cite{nowak} concentrates on the characterization
(Chomsky hierarchy) of languages and the learning of languages. One section
simulates the evolutionary competition of languages or grammars,
when one generation learns it from the preceding one. Its computer simulation,
Fig.4 of \cite{nowak}, gives a transition between dominance of one learned 
language and the coexistence of many languages with similar frequencies. Only
in the first case of dominance a child has learned a working language. 
More information on simulations of language evolution is given in \cite{cang}.

This model has $n$ languages $L_i, \; i=1,2.\dots L$. $F_{ij}$ is the pay-off
(advantage to be understood) for a speaker of language $L_i$ talking to a 
speaker of language $L_j$; the fitness of $L_i$ thus is $f_i = \sum_j F_{ij}x_j$
and depends on the set of fractions $x_j$ of the population which speak $L_j$.
The mutation probability that children from $L_i$-speaking parents will speak 
$L_j$ is $Q_{ij}$. (Komarova \cite{komarova} adds a $x$-independent term 
to this $f_i$ and allows the $Q_{ij}$ to depend on the $x_i$.) 
The average fitness is $\phi = \sum_i f_ix_i$. Then their
differential equation is 
$$dx_j/dt = (\sum_i f_iQ_{ij}x_i) - \phi x_j$$
and describes a survival of the fittest as in biology. Moreover, if all 
languages increase their fitness by the same amount, the RHS of this equation 
will not change, as in known from the nuclear arms race in human history, or
the Red Queen syndrome in biology. Instead, a high learnability $Q_{ii} \simeq
1$ and high ability $F_{ii}$ of communication helps language $L_i$ to win: 
$x_i = 1, \; f_i = \phi, \; dx_i/dt = 0$ is the fixed point where everybody 
speaks $L_i$.   
 
For a large number $L$ of languages, one has a huge number of free parameters
in the $L \times L$ matrices $F_{ij}$ and $Q_{ij}$. For only $L = 3$ languages,
suitable choices for the $Q_{ij}$ allow for limit cycles instead of
fixed points as solutions \cite{mitchener}; for $L=5$ also chaos is possible.

The motivation for these differential equations is not the competition of
languages spoken by adults but the learning of a language by children; see
also \cite{komarova} and further literature cited there. But perhaps the same
mathematics can be applied for language competition as reviewed here.
Simulations how the first proto-language may have evolved for pre-linguistic
humans were presented already earlier \cite{krakauer}.

\bigskip
Nettle \cite{comrie} tried to explain why in the only recently
populated Americas one has a larger language diversity than 
in the older Africa or Eurasia. With a low population density,
little competition between languages exists, but for longer times
$t$ (measured in thousand years) the resulting growth of diversity
is diminished, proportional to $1/t$. Also, extinctions happen 
with a probability of five percent per millenium. Then \cite{comrie}
$$dL/dt = 70/t - L/20$$
gives a number of languages $L$ first increasing steeply, having 
a maximum and then decaying slowly. Later, we will bring in
section 4.2.2 a more microscopic justification, which moreover
avoids the complete extinction of $L$ for very long times
inherent in this Nettle equation. (Actually, he uses discrete 
time and his $L$ is the number of language groups, called
stocks.) 

A treatment of the model of the next section with differential equations,
is in preparation by Saakian \cite{saakian}. 

\subsection{Many Languages: Probability Distributions}

Pagel \cite{pagel} assumes the total number of languages to increase
with time $t$ as exp$[(\lambda - \mu)t]$.  Each language changes with
a rate $r$ such that two groups of people, initially sharing one language, 
have only a fraction  exp($-rt$) of language in common after a time $t$ has
elapsed; $r$ is a few \% per century for human languages. The 95 Indo-European
languages differ much less in their rates $r$ than the 371 Malayo-Polynesian
languages, as described by suitable probability distributions of $r$ values. 
Moreover, important words in one language may change less than unimportant 
words. He mentions that the rate of language extinction is two to eight 
times higher than the production rate of new languages, worldwide, and that
perhaps only five hundred languages will survive the twenty-first century.
Such effects will be simulated below, Fig.4. He also reviews the human
language diversity, similar to mammalian species diversity \cite{sutherland},
as a function of geographical latitude in North America; such effects could
be studied by the methods of \cite{finland} or by putting the microscopic 
model of the now following section onto an inhomogeneous lattice.

\section{Microscopic Simulations}

Most of the microscopic simulation known to us are our own \cite{schulze}, and 
thus we review them and add additional later simulations not yet published
by us. But we mention here that an earlier paper of Briscoe \cite{briscoe}
contained simulations with a more complicated model, using only $10^2$ people. 
The aim of that paper was grammar acquisition, how children learn to speak
a language properly, and this is outside our area of competition between
different languages for adults. Nevertheless Fig.13 there looks like 
language competition, and the long paper contains lots of useful background
information. Before we come to our model, we review in the next subsection
a microscopic two-language model \cite{kosmidis}.

\subsection{Two languages}

Kosmidis et al \cite{kosmidis} follow \cite{strogatz,finland} by studying only
two languages, but they simulate individuals which are born and die as in 
\cite{schulze}. Moreover they put them onto a lattice. Each person has a 
bit-string of 20 ``words''. Initially, those who speak the first language have
their first 10 bits set to one and the other 10 bits set to zero, while for
the people speaking the other language
the first 10 bits are zero and the other 10 bits are one.
Later everybody can increase its biological fitness (probability to reproduce)
by learning words of the other language, changing zero to one, without 
necessarily forgetting its original language. Thus synonyms become possible.

People diffuse on a dilute 
$100 \times 100$ square lattice and may learn words from 
others when they bump into each other. In the case of no births and no deaths 
the populations end up speaking on average five words from one and five words
from the other language. With birth and death included, people increase their
vocabulary until everybody speaks nearly all 20 words, i.e. is completely
bilingual. Only the emergence of short-range order, not that of long-range
order, is published in \cite{kosmidis}. The model has five independent 
probabilities as free parameters, plus the two initial concentrations and
fitnesses.

What are called ``words'' by \cite{kosmidis} can also be grammatical principles,
% This paragraph added
like whether a normal sentence is ordered subject-verb-object or subject-object-verb
\cite{science}. Or it can be the shape of a letter for an alphabet. These are
details which we ignore, just like Kepler ignored the well-known extension of
the Earth. And the same remark applies for our following model.

\subsection{Our Basic Model}

\subsubsection{Model Definition}

Languages are described by strings of $\ell$ bits, each of which can be zero or 
one; $\ell$ is fixed for each simulation to a value between 8 and 64. Such a 
bit-string was already used by \cite{nowak,briscoe} for languages. In contrast 
to their work and to biology we assume in the basic model that the intrinsic 
usefulness or ``fitness'' of all languages is the same. Instead, languages of 
small size have a tendency to shrink and to go extinct because speakers switch
from the small language to one of the big ones. In reality, humans sometimes 
cling to their small language as part of their identity, an effect which still 
needs to be simulated. A set of 30 independent binary grammatical parameters,
i.e. $\ell = 30$, was regarded as reasonable by Briscoe \cite{briscoe}.

We start with one individual speaking language zero (all bits set to 0). Then,
at each iteration $t = 1,2,\dots$, each living individual gives birth to 
one child. The child speaks the same language as the parent except that 
with probability $p$ one of the $\ell$ bits is randomly toggled, i.e. switched 
from zero to one, or from one to zero. Note that $p$ is the probability per
individual; the probability per bit is $p/\ell$. If for real human languages,
words change at a rate of $p/\ell \simeq  2$ \% per century, our typical values
$p = 0.48, \; \ell = 16$ require several centuries for one iteration 
\cite{pagel}.

If we would not include any deaths 
the population would double at each iteration. Thus we introduce for each
individual a Verhulst probability $= N(t)/K$ to die from starvation or 
lack of space, where $N(t)$ is the total population at the beginning of time
step $t$; often $K$ is called the carrying capacity. Now the population 
doubles at first for each iteration, until it reaches a plateau at $K/2$.

In this version one has now a stable population but eventually all languages 
will appear with the same probability, hardly a realistic result. Languages
are ways of communication, and at least for spoken languages, as opposed
to alphabets for writing, usually one person speaks to another. Thus languages
spoken by only one person are less useful than those spoken by many people. 
Therefore we assume that a speaker of language $i$ of size $S_i = x_iN(t)$
switches with a probability 
proportional to $(1-x_i)^2$ to the language of a randomly selected person
in the whole population, which usually will be a language of large size. The
exponent two should take into account that the usefulness of a language $i$
is proportional to the square of its concentration $x_i$, if two people talk;
other choices like $1-x_i^2$ have also been tried.
  
This pressure to switch to widespread languages is assumed to be strong for
high population densities (``globalization''), and not at the beginning 
when $N(t) \ll K$. Thus we take the switching probability $r$ also proportional
to $N(t)$, and therefore the complete switching probability is 
$$ r = (2N(t)/K)(1-x_i)^2 \quad \rightarrow \quad (1-x_i)^2 \quad 
{\rm for} \quad t \rightarrow \infty \quad .$$ 

Instead of starting with one person speaking language zero, we also started 
with the equilibrium number $K/2$ of people and let them initially either all 
speak language zero, or different languages randomly selected for each 
individual.

This finishes the definition of our basic model

\subsubsection{Results}

The $P(S)$ of real languages \cite{sutherland,science} is roughly a log-normal 
distribution with enhanced numbers for small sizes $S$, Fig.2a. Such behaviour
is seen also in the simulated curve of Fig.2b. 
Fig.3 shows more systematically the simulated distribution $P(S)$ of language 
sizes, binned by powers of two thus that the upper end of the interval is twice
the lower end, like all sizes between 32 and 64 put together.
Real languages are not in equilibrium, and thus, in contrast to typical 
physics simulations, the histograms in Figs.2,3 are not taken in an
equilibrium situation but averaged over all 1000 iterations. (A simulation
similar to Fig.2b but using only $500 < t < 1000$ iterations gave nearly
% test added
the same results as $0 < t < 1000$, apart from a reduction by a factor two.)
To find phase transitions in what follows, we wait a few hundred time steps to
see what happens.

Fig.4 shows the increase of the population up to the plateau at $K/2$.
The final population is $K/2$ and not $K$ since we determine the Verhulst 
probability $y = N(t-1)/K$ at the beginning of iteration $t$ and leave it at
that value for the whole iteration. The Verhulst deaths thus reduce the 
population by a factor $1-y$, and if each of the survivors has $b$ offspring,
the population is multiplied by another factor $1+b$. For a stationary
population, these two factors have to cancel: $(1-y)(1+b) = 1$, giving
$y = b/(1+b) = 1/2$ for our choice $b=1$.

As a function of $p$, Fig.5 gives the size of the largest language.
Ten samples were simulated for each mutation rate and their
results are shown besides each other for proper visibility; thus the horizontal
axis is actually the line number of the output and roughly a linear function 
of the mutation rate $p$ which varies here between 0.240 and 0.328 in steps of
0.008. 

In Figs. 3 and 5,
varying the mutation rate $p$ we see two different regimes: For small $p$
one language dominates (usually it is language zero with which we start) and
is spoken by nearly everybody; for large $p$ the size distribution $P(S)$ 
is nearly log-normal, and no language is spoken by far more people than 
the others. The first choice, ``dominance'', may correspond to the present
use of alphabets, the second one, ``fragmentation'', to that of human 
languages. As mentioned above, also Nowak et al \cite{nowak} distinguish
dominance (useful for learning a language) from a multitude of languages with
roughly equal population sizes. 
Fig.5 starts with one person and then see some hysteresis: Some
of the samples show dominance of one language, and others not. The width of 
this transition region is roughly independent of the population size. 

As Fig.6 shows, the position (as opposed to the width) 
of this transition depends strongly on the 
population size and also on how we start: one language or many languages?
Clearly, if we start already with everybody speaking one language, dominance
is easier to maintain and requires a higher mutation rate to be broken.

Dominance is not unrealistic:
The hieroglyphs of Egypt were an alphabet but have gone out of fashion. 
This paper uses the Roman alphabet, which together with the Hebrew, Greek,
Arabic, Cyrillic, ...alphabets originated from one invented more than three
millennia ago on the eastern shores of the Mediterranean. Thus the Roman 
alphabet is now clearly dominating together with the other mutants; Chinese
writing is not an alphabet.

If one language dominates, it is not necessarily the zero if we start with one 
speaker. Some mutations in the very first iterations may have caused a mutant
to get ahead of the original zero and to win finally, just like the Roman 
alphabet was not the first of the above-mentioned family. 

Physicists like scaling laws, and indeed a rather simple scaling law is found
here. If the possible number $M = 2^\ell$ of languages is much smaller than the 
population size $N = K/2$, then every language will be spoken by some people. 
In the opposite limit of many more possible languages than people, not 
everybody will have his/her own language since then communication would be 
impossible. Thus the people cluster into small groups of one language each,
with the actual number $L$ of languages proportional to the number $N$ of 
people. In between, a simple scaling law holds, as shown in Fig.7:
$$L/M = f(M/N) \quad .$$
And this scaling function $f$ is independent of the length $\ell$ of the 
bit-string for 8 and 16 bits. (For 32 and 64 bits, $M$ is so large than only
the right tail outside Fig.7 was simulated: $L/N \simeq 0.016$.) As $\ell=30$
was mentioned by Briscoe \cite{briscoe}, meaning $M = 10^9$ possibilities,
a human population of $10^9 \dots 10^{10}$ people would fit into the transition
region between the two straight lines of Fig. 7.

Actually, this scaling law was seen before in opinion dynamics \cite{sousa}
where the number of surviving opinions depends on the number of possible 
opinions and the number of people. Our switching from a small language
to one of a randomly selected person is similar to the opinion dynamics of 
Axelrod \cite{axelrod}, and we may also interpret the $\ell$ bits in our
bit-string as $\ell$ different yes-no opinions on $\ell$ different themes
\cite{jacob}. (In turn, the Latan\'e opinion dynamics model  was applied to 
languages in \cite{nettle}.)

An analysis of the Hamming distances (similarities and differences) between
the different simulated languages is planned \cite{hmo}. The lifetimes of
languages are distributed exponentially except for short lifetimes 
\cite{schulze}.  Now we explain modifications of this basic model.

\subsection {Approaching Languages}

\subsubsection{Imitation}

Refs.\cite{strogatz,finland} used already the concept of one language being 
better than another. For example, many French words were taken over into
German language because of the then advanced French civilization; 
names of beers  travelled in the opposite direction. Thus we assume that
with an imitation probability $q$ the mutation at the randomly selected position
during the birth of a child does not toggle the bit but takes it from 
the superior language, which has all bits except one set equal to 1. 
For $p=0.48$ which for both lengths $\ell = 8$ and 16 avoided dominance
in the basic model, one now finds the superior language to become the largest
one, except for small $q$ (not shown). 
For imitation probabilities $q$ larger some threshold increasing with population
size and decreasing with bit-string length, the superior
language becomes the strongest in more than half of the ten samples simulated,
but it is not yet dominating in the sense of being used by nearly all people. 
For even larger $q$, this strongest superior
language may even become dominating and then is spoken by nearly everybody. 
In some sense the concept of a superior language here corresponds to the
superior status $s$ of the Abrams-Strogatz model.

\subsubsection{Merging Languages}
Differently from biology, two different languages can become more 
similar to each other and finally merge into one, like English
being a merger of French (Norman) and German (Anglo-Saxon) \cite{bragg}.
Thus in the present modification we start from all languages equal, 
and then with a transfer probability $q$ a mutation does not flip a bit
but takes a bit from the language of another randomly selected
person. The Verhulst factor applies as usual, and so does the 
flight from small languages. We start with the equilibrium size
of the population and many randomly selected languages. 
We thus have in addition to the fixed switching probability $r$ and the
% clarification added
adjustable mutation probability $p$ also an adjustable mutation 
probability $q$. 

In this modification, depending on the mutation 
probability $p$ (favouring language diversity) and transfer
probability $q$ (favouring dominance of the largest language),
one finds either one language comprising nearly all people,
or many roughly equally large languages. Fig. 8 shows our
phase diagram for $10^5$ people. In the upper left part of high
transfer and low mutation probability, nearly everybody at the 
end speaks the same language. In the lower right part of low 
transfer and high mutation probability, language diversity 
persists. We now see, in contrast to our earlier results, a 
strong difference between 8 and 16 bits: Dominance is rare for 
16 bits. 

In Fig.8a we started with a large number of people speaking randomly selected 
languages. Starting instead with one person only, and using the modified 
program allowing up to 64 bits, we do not see in the phase diagrams of Figs.8b,c
the strong dependence on the number of bits evident from Fig.8a. As already
Fig.6 made clear, the position of the phase transition depends strongly
on the initial conditions; now we see that initial conditions also influence 
whether or not the length $\ell$ of the bit-string is important. 
Fig.1 of \cite{komarova} and Fig.7.3 of 
Komarova and Nowak in \cite{cang} give a similar phase
diagram for the differential equations of section 3.2. 

The case of English as a merger of two languages may be better 
described by a simulation starting with only two languages,
each having half of the equilibrium population. Transfer
and imitation are neglected: $q=0$. For small mutation rates $p < 0.55$,
one of the two languages wins and is spoken by nearly everybody;
for larger $p$ no dominance appears and the two languages are just two of many. 
The threshold for $p$ for $\ell = 8$ is roughly size-independent and near 
0.52, not shown.  The short time for equilibration allows simulations of 1000
million people, Fig.9.  If eventually one language dominates, it is
usually one of the two starting ones and not a mixture, in contrast to
English; a better mixture model is discussed in Refs.23, 21b.
A somewhat analogous probabilistic model for 
the {\em learning} of two possible language grammars from presented
correct sentences was published much earlier by Niyogi and Berwick
\cite{berwick}; see also \cite{kazakov}.

\subsubsection{Size and Time Effects}

Fig.10 shows strong size effects for the position of the phase
transition: At a fixed mutation probability $p = 0.48$, the
minimal transfer probability $q_c$ needed for dominance of one
language strongly increases with increasing population size 
$N$ and may vary as $1 - {\rm const}/\log(N)$ for $N \rightarrow
\infty$. We offer the following argument to explain this
size effect:

All languages start as equals, but due to random fluctuations
after some time one language has such a strong advantage over 
the other languages, that this advantage grows further due to the
flight from small languages, until at the end nearly everybody
speaks this dominating language.

If correct, then the time development should consist of a
random initial part of fluctuating length, followed by a 
deterministic growth of the largest language. Biologically, 
HIV infection might behave similarly. Indeed, Fig.11 for $10^8$
people shows nicely the deterministic and roughly exponential 
growth of the second part. Fig.12 shows that the larger
the population is, the longer is the initial part where no
language has a very clear advantage; indeed, for larger $N$ 
the relative fluctuations become smaller and a random victory 
for one language takes more time. In Fig.13 we see from 1000
samples that the total times after which dominance of one 
language is achieved, fluctuate strongly and may follow a 
log-normal distribution. Its average increases roughly with the
square-root of the population, while the logarithmic width decreases
slowly for increasing populations.

More explicitly, we tested our hypothesis by giving language 
zero an initial advantage of 5 percent of the total population
and then check whether this favoured language zero, or another 
language, wins at the end. As Fig. 14 shows, for small
populations another language can win (part a) since fluctuations are
relatively strong. On the other hand, for large 
populations the favoured language wins (part b) since the initial
advantage is not masked by fluctuations. Fig. 15 shows more
quantitatively the probability of the initial leader to win at 
the end, as a function of $N$ (from 1000 and 100 samples).

Thus these tests are compatible with our above explanation;
it is therefore not clear if our phase transitions would persist
for infinitely large populations observed over long but finite 
times.

Fig.16 shows clearly that the times for dominance of one 
language vary with $q$. Is it possible that for infinite times
always dominance would appear? Fig.17 answers this question with
no: The inverse logarithm of the time to reach dominance 
decreases with decreasing $q$ and seems to vanish at some
positive $q_c$, suggesting a sharp phase transition and not a
gradual freezing in; freezing-in would correspond to $q_c=0$ and is difficult
to reconcile with extrapolations of Fig.17. 
Thus while our above phase diagram, based
on 1000 iterations, may shift if we increase the observation
time, the phase transition would not vanish for $t \rightarrow
\infty$ at a fixed large population size.

\subsection{Interface Structure}

When two ethnic communities live in one city, sometimes 
ghettos are formed, i.e. homogeneous regions with one community
are adjacent to homogeneous regions of the 
other community. Since the classic paper
of Schelling \cite{schelling} this effect was simulated
with Ising-like models \cite{ortmanns}, also for more than 
two communities \cite{schulzepotts}. A famous example was New Orleans,
where Canal Street separated the French Quarter in the West from
the English regions eastward. We now want to study linguistic 
geographical separation, similar to \cite{finland}.

In Ising models such simulations have been made since decades
with initially all spins up in one part and all spins down in 
the other part of the system. (In these magnetic models, a 
spin is a variable which is either up or down, representing 
the magnetic orientation of the atom.) It is not necessary to
employ the bias of \cite{finland} where as in a magnetic field
one orientation was preferred in one part and the other in the
other part of the system. For large enough systems, the 
``inertia'' suffices to keep the two orientations stable 
over long times. (See Livingstone versus Nettle for a similar
discussion on whether one language should be favoured in simulations
\cite{cang}.)

Thus we put our people onto a $10 \times 10$ square lattice,
with antiperiodic boundary conditions \cite{landau}. (For periodic
boundary conditions, the left neighbour of a site at the left lattice
% explanation added
boundary speaks the same language as in the same line the right-most
site; for antiperiodic boundary conditions the left neighbour speaks the
compliment of the language at the right-most site, i.e. all bits are toggled.)
On each of the 100 lattice sites the standard language
competition is simulated, with the following modifications:
If a speaker of a small languages switches to a larger language,
then the language of a randomly selected person on the same 
lattice site is taken. On the other hand, if a speaker 
during a mutation with probability $q$ selects a random bit from
the language of another person, then this person is selected
in half of the cases from the same lattice site, and otherwise
from one of the four nearest neighbour sites. In addition, with
a low probability $d$, at each iteration every person exchanges
places with a randomly selected other person from one of the four
nearest neighbours, as in Kawasaki spin-exchange dynamics 
(diffusion). Thus, in contrast to \cite{schelling}, we do not 
need empty spaces to facilitate this motion of people within
their city. (Also, instead of the complicated formula at the end of section
4.2.1, we use $1-x_i^2$ as switching probability $r$.)

In this model, we start with the equilibrium number of people
such that in the left part each person speaks language A =
00000000 while in the right part everybody speaks B = 
11111111; all languages contain 8 bits only. 

If every site carries on average 50 people, thus corresponding to
a large apartment building, then no site is completely emptied
in our simulations, with only 25 people a few sites became empty. 
However, the interface between A and B dissolves
after a few hundred iterations: inertia is not big enough. With
100,000 people per lattice site stable reproducible results are
found instead. 
 
For $d=0$ the interface remains very sharp: the population of
A speakers jumps from nearly 90 percent to exactly zero 
within one lattice constant. For $d = 0.01$, on the other hand,
a more interesting interface structure is found, Fig. 18,
where the fraction of A  speakers in the part dominated by
language B decays exponentially with distance from
the separation line, similar to tunneling of electrons in
quantum physics. %In contrast to the up-down symmetric Ising 
%interface, our model has a much stronger deviation from  the 
%bulk concentration of language A in the A region, than the
%deviation from zero concentration in the B region. By 
%construction, A and B behave symmetrically to each other. 
Increasing the lattice size does not increase the interface width\cite{schulze}.

The interface may also serve as a heterogeneous nucleation centre causing 
the metastable state to decay, just like water vapour in the winter may condense
on cold windows but not in the free air. Thus if we start with a homogeneous 
state of one language only, and compare it with an initial coexistence of two 
languages (one in the right part, one in the left part), which state decays more
easily? For large transfer probabilities $q$ near 1, the initial dominance or
coexistence is preserved at intermediate mutation probabilities $p$, 
in both cases. But already at a slight reduction in $q$, the two-phase 
coexistence decays (usually into fragmentation), while the homogeneous initial
state needs a much stronger reduction of $q$ to fragment, Fig. 19.
% following remark added
(Fig.8 shows that we have dominance for $q=1$, which should be compared with the
homogeneous bulk phase of nucleation physics. Thus the difference $1-q$ is the
perturbation leading to nucleation, and this perturbation can be much smaller
if the system is already perturbed by an interface allowing heterogeneous 
nucleation.)

\section{Summary}
Computer simulation of language competition is in its infancy, and several
models have been tried. Future simulations should clarify the similarities
and differences for the various approaches. For example, can an analog of the 
Zipf law or its recent variants \cite{gonc} be studied? Can 
a population stick to a rather small language which they regard as important for
their identity, like French in North America? Or could one simulate people 
speaking several languages?

The microscopic method, which not only we \cite{briscoe,kosmidis}
prefer, shows a phase transition where the fraction of people speaking the
largest language jumps from a low value to more than 1/2. The position of 
this transition in the plane of the mutation probability $q$ and transfer 
probability $q$ depends also on the initial conditions: Do we start with
one or with many languages? It also depends strongly on the population size
and waiting time, 
and sometimes on the length of the bit-strings used to represent a language.

It would be helpful if linguists would point out the properties of language
size distributions and other aspects of reality which they regard as important 
and which such simulations could reproduce. 

DS thanks the German-Israeli Foundation for a travel to Israel where he 
gave his first talk on these simulations, P.M.C. de Oliveira for suggesting 
to simulate languages, and J. Adler, E. Briscoe, F. Schweitzer, M. Pagel and 
D.B. Saakian for helpful comments.

\bigskip
\bigskip
{\bf \large Appendix: Computer Program}
\bigskip

To encourage readers to start their own simulation we give here our Fortran 
program for section 4.1, with explanations. Experts will notice the similarity 
with the Penna model of biology \cite{penna,pmco}; for example the bit-string
is called gen1f as if it would be a female genome.

\bigskip

{\small
\begin{verbatim}
      implicit none
      integer popdim,nbyte,nbit,nshift,len,irun,nrun,nhist(1000000)
      real fmut,rand,select
      parameter(nbyte=1,popdim=20000    ,nbit=8*nbyte,len=2**nbit)
c             integer*2 gen1f(popdim), gene1, p, bit(0:nbit)
              byte      gen1f(popdim), gene1, p, bit(0:nbit)
      integer popmax,inipop,maxstep,fage,k,nlog(0:30),j,
     1 birth,t,i,seed,fa,n,fpop,nlabel(-len:len),number,kmut
      parameter(popmax=popdim,inipop=1,maxstep=1000,nrun=1  ,
     1 fage=nbit,birth=1,seed=1)
      integer*8 ibm,verhu,mult,imut
      nshift=0
      if(nbyte.eq.2) nshift=60
      if(nbyte.eq.1) nshift=61
      if(nshift.eq.0) stop 9
      ibm=2*seed-1
      mult=13**7
      mult=mult*13**6
      fmut=rand(seed)
      print *, popmax,inipop,maxstep,
     1         fage,birth,seed,nbit,nrun
      do 18 kmut= 600,600,100
      fmut=kmut*nbit*0.0001
      print *, fmut,kmut
      if(fmut.ge.1.0) stop 9
      imut=2147483648.0d0*(fmut*4.0d0-2.0d0)*2147483648.0d0
      bit(0)=1
      do 2 i=0,nbit
        if(i.gt.0) bit(i)=ishft(bit(i-1),1)
 2      ibm=ibm*16807
      do 15 k=1,1000000
 15     nhist(k)=0
      do 11 irun=1,nrun
      fpop=inipop
      do 6 i=1,fpop
 6      gen1f(i)=0
      select=2.0/popmax
c     
      do 7 t=1,maxstep
      verhu=2147483648.0d0*(fpop*4.0/popmax-2.0)*2147483648.0d0
      do 3 i=-len,len
 3     nlabel(i)=0
      do 4 i=1,fpop
 4      nlabel(gen1f(i))=nlabel(gen1f(i))+1
      if(t.eq.(t/100)*100) then
        number=0
        do 5 i=-len,len
 5        if(nlabel(i).ge.10) number=number+1
        print 8,irun,t,fpop,number,nlabel(0),(nlabel(2**i),i=0,4)
 8      format(2i5,3i10,5i8)
      end if
      i=1
      fa=fpop
 9    if(rand(0).lt.(fpop*(1.0-nlabel(gen1f(i))*1.0/fpop)**2)
     1   *select) then
 14     k=1+rand(0)*fpop
        if(k.le.0.or.k.gt.fpop) goto 14
        gen1f(i)=gen1f(k)
      end if
      ibm=ibm*16807
      if(ibm.lt.verhu) then
c     death 
        if(fpop.le.1) goto 1 
        gen1f(i)=gen1f(fpop)
        fpop=fpop-1
        if(fpop.ge.fa) then
          i=i+1
        else
          fa=fa-1
        endif
      else
c     survival
        do 12 n=1,birth
          gene1=gen1f(i)
          fpop=fpop+1
          if(fpop.gt.popdim) goto 1 
          ibm=ibm*mult
          if(ibm.gt.imut) goto 13
c         Exactly one mutation is made with probability fmut
          ibm=ibm*16807
          p=bit(ishft(ibm,-nshift))
          gene1=ieor(gene1,p)
 13       continue
          gen1f(fpop)=gene1
 12     continue
        i=i+1
      endif  
c     if(death) .. else (survival, birth) ..
      if(i.le.fa) goto 9
 7    continue
      do 10 i=-len,len
        if(nlabel(i).eq.0) goto 10
        k=min0(1000000,nlabel(i))
        nhist(k)=nhist(k)+1
c       if(irun.eq.nrun) print *,i,nlabel(i)
 10   continue
 11   continue
      do 19 k=0,30
 19     nlog(k)=0
      do 16 j=1,1000000
      k=1.0+alog(float(j))/0.69315
 16   nlog(k)=nlog(k)+nhist(j)
      do 17 k=0,30
 17     if(nlog(k).gt.0) print *, 0.707*2**k, nlog(k)
 18   continue
      stop  
 1    print *, 'error',fpop 
      end
\end{verbatim}
}
\bigskip

The program allows for bit-strings of length $\ell = 8$ stored in words of 
type {\tt byte}, or of length $\ell = 16$ using type {\tt integer*2}. This
choice has to be made in the {\tt parameter} line and the line following it.
We have $2^\ell$ possible languages, each of which can be stored easily. For
$\ell = 32$ and 64 we used a different, more time consuming program available
from us as {\tt language20.f}.

Our random numbers are 64-bit integers {\tt ibm} with $-2^{63} < {\tt ibm} < 2^{63}$
produced by multiplication with 16807 or with
{\tt mult }$ = 13^{13}$; in addition
we use a built-in random number generator rand(0) to give real numbers between
0 and 1. If only 32-bit integers are available, readers have to adjust the
lines where $2147483648 = 2^{31}$ appears.

Loop 7 is the main time loop, and we now describe in the order of the 
program   what happens at each iteration. Loops 3 and 4 count in
{\tt nlabel(.)} how many individuals speak a given language. Every 100
time steps the total {\tt number} of languages spoken by at least ten 
people is determined and printed out together with some of the language
sizes {\tt nlabel}.  

The 37 lines following {\tt i=1}, upto {\tt goto 9}, are the loop over all
individuals $i = 1,2, \dots,$ {\tt fpop}. Since the population {\tt fpop}
varies due to the death and birth processes, we did not deal with them
in a fixed loop {\tt i=1,fpop}, and instead used the backward jump 
{\tt goto 9} and the number {\tt fa} of adult individuals; at the beginning
of the iteration all individuals are adult: {\tt fa = fpop}.
 
The six lines starting with label 9 simulate the switching from a rare 
language to that of a randomly selected individual {\tt k}. Then comes
an {\tt if then else endif} choice between Verhulst death and survival.
In case of death, the last individual {\tt fpop} is put into the place 
of the now dead individual {\tt i}, and if {\tt fpop} was a child born
during the same iteration, then the counter {\tt i} for the individual 
is increased by one since this child is not subject now to Verhulst deaths.
Otherwise the number {\tt fa} of adults to be treated decreases by one.  

In the case of survival instead of death, the counter {\tt i} always increases
by one, and loop 12 allows for the birth of several children. Each child
increases {\tt fpop} by one, gets a random bit position {\tt ishft(ibm,-nshift)}
between 1 and $\ell$, and has the bit at that position changed with an exclusive-or
command {\tt ieor}. 

After the {\tt if then else endif} choice between death and survival, 
we jump back to label 9 if the counter {\tt i} is not larger than the
number {\tt fa} of adults to be treated. In this way all the adults, including
the ones which replace the dead ones, are treated once, while the children
born during this iteration neither die nor give birth.

One could add an age structure for the adults, with reproduction starting
at a minimum age only  and a genetic mortality increasing exponentially 
with age \cite{penna,kunwar}; this was not yet done by us.

\begin{figure}[hbt]
\begin{center}
\includegraphics[angle=-90,scale=0.4]{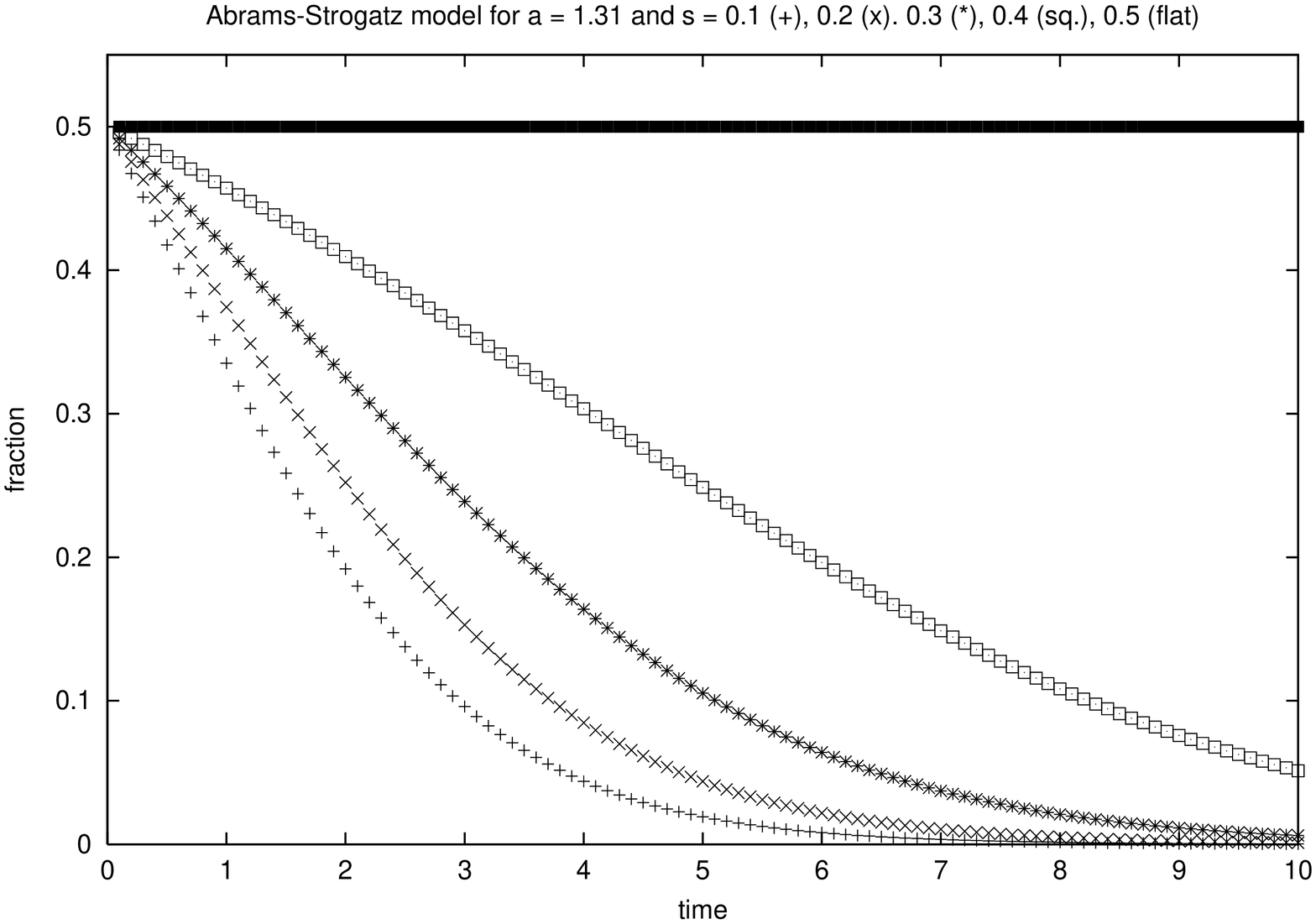}
\includegraphics[angle=-90,scale=0.4]{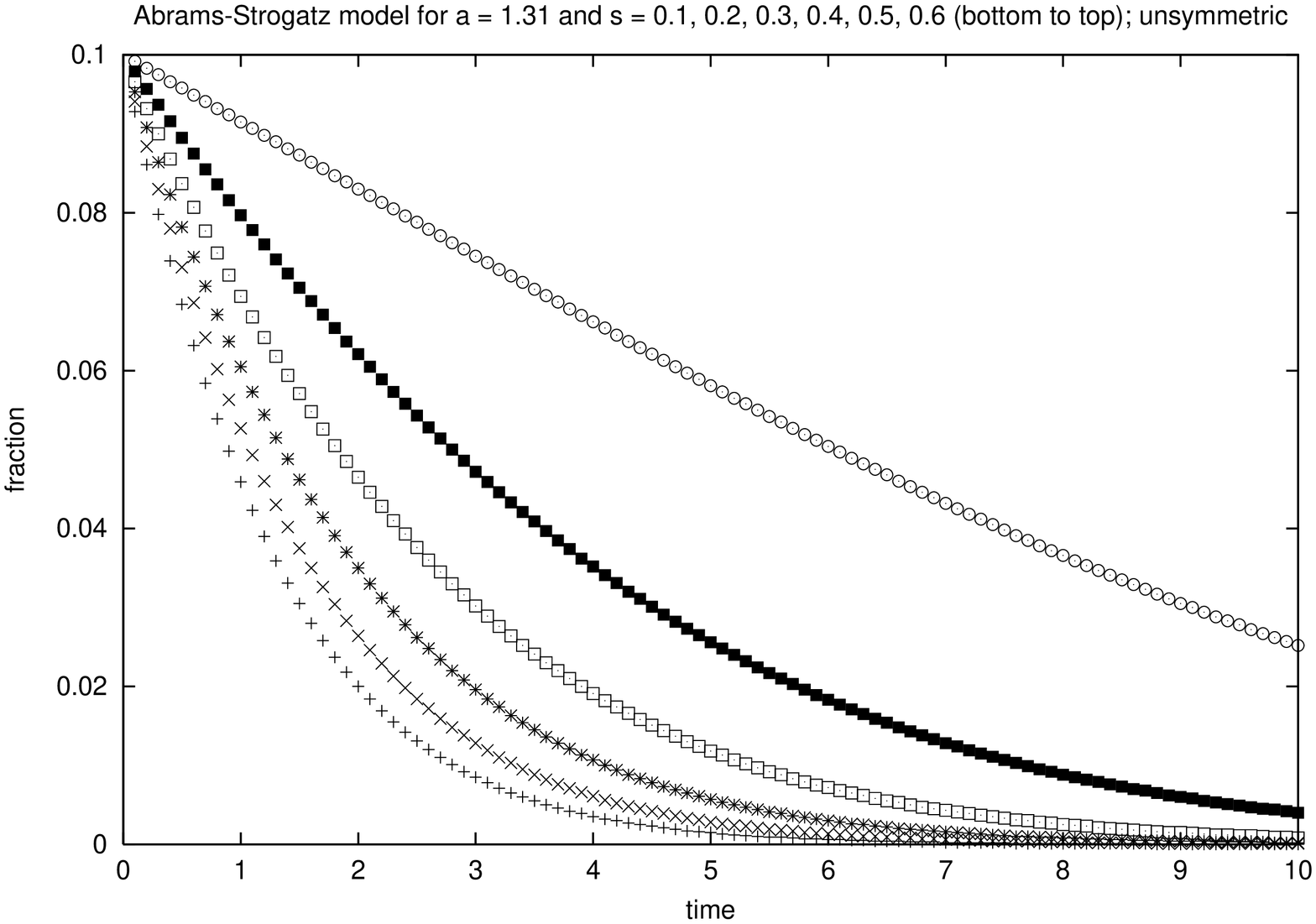}
\end{center}
\caption{
Fraction of population speaking first language in the two-language model of
Abrams and Strogatz \cite{strogatz}.
Part a: Initial distribution symmetric: Each language is spoken by half the
population. Now the one with the lower status $s < 1/2$ goes to zero. (The
symmetric curves for $s> 1/2$ aopproach unity and are not shown.) Part b:
Initial distribution favours second language; now first language dies out 
even for $s = 0.5$ and 0.6.
}
\end{figure}
\begin{figure}[hbt]
\begin{center}
\includegraphics[angle=-90,scale=0.4]{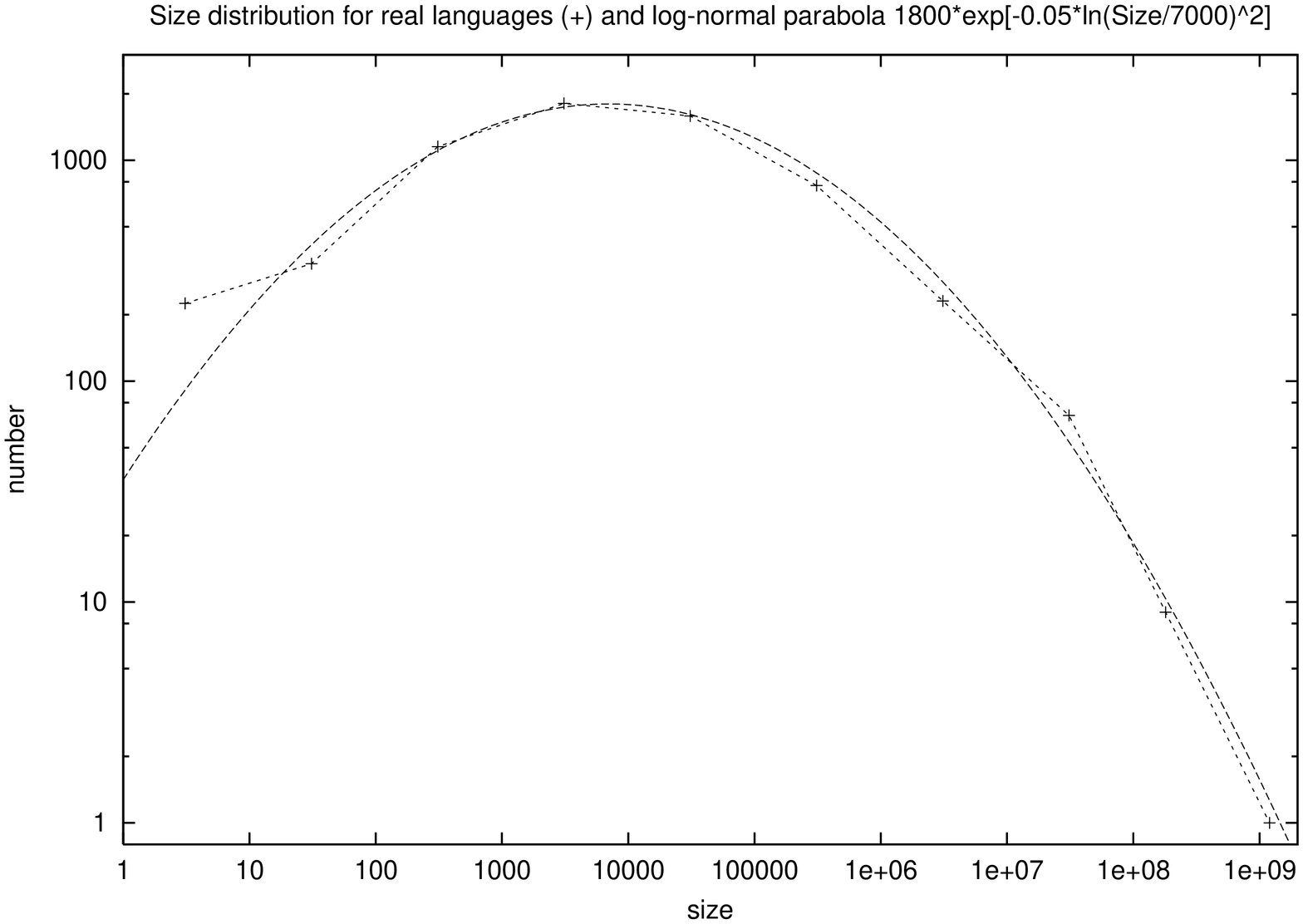}
\includegraphics[angle=-90,scale=0.4]{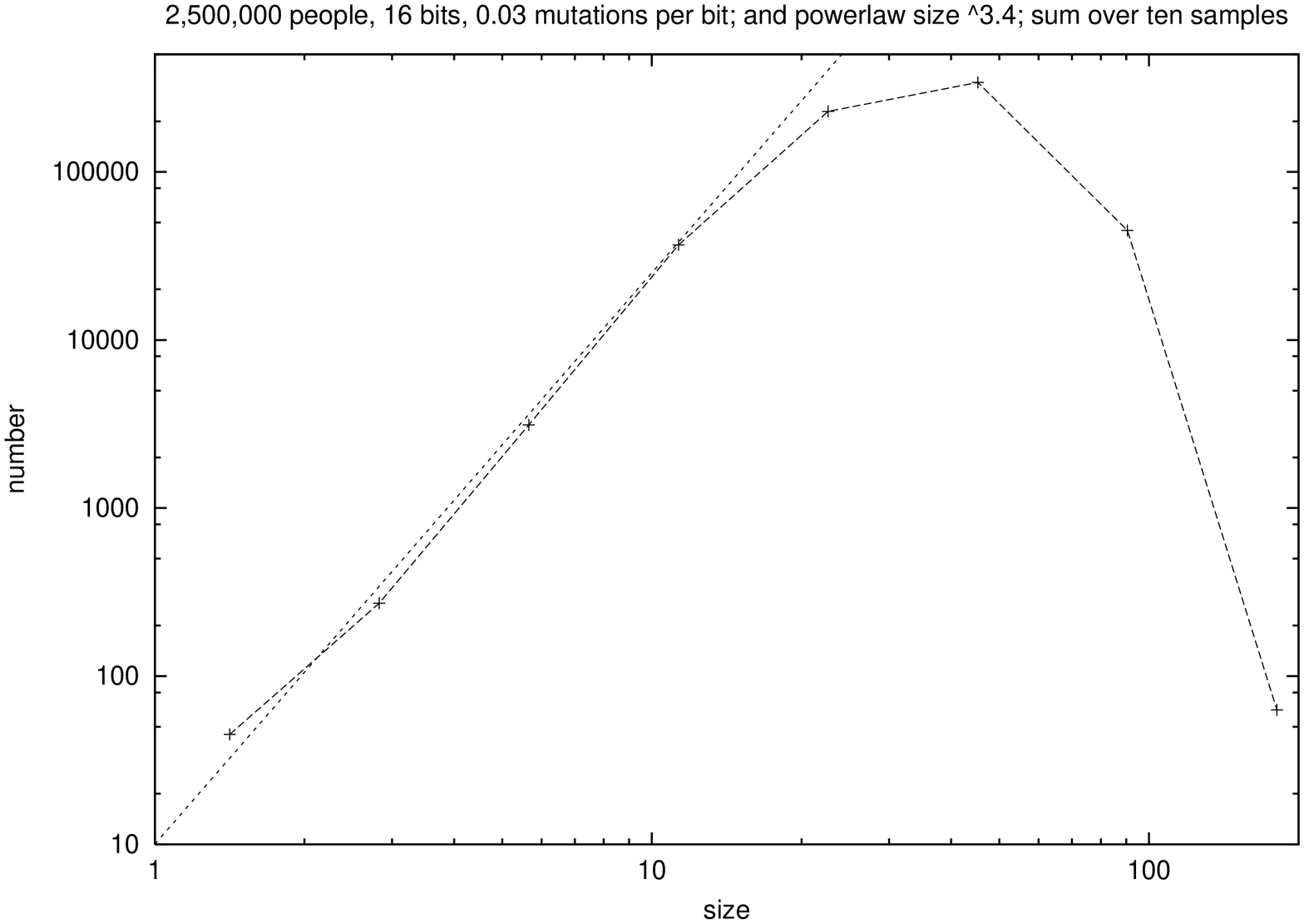}
\end{center}
\caption{
Part a: Present real distribution $P(S)$ of human languages; the parabola 
corresponds to a log-normal distribution which is disobeyed at the left end.
Part b: Simulated histogram $P(S)$ (from \cite{schulze}); the straight line 
corresponds to a power law. We see qualitative but not quantitative agreement
between reality (top) and simulation (bottom).
}
\end{figure}

\begin{figure}[hbt]
\begin{center}
\includegraphics[angle=-90,scale=0.40]{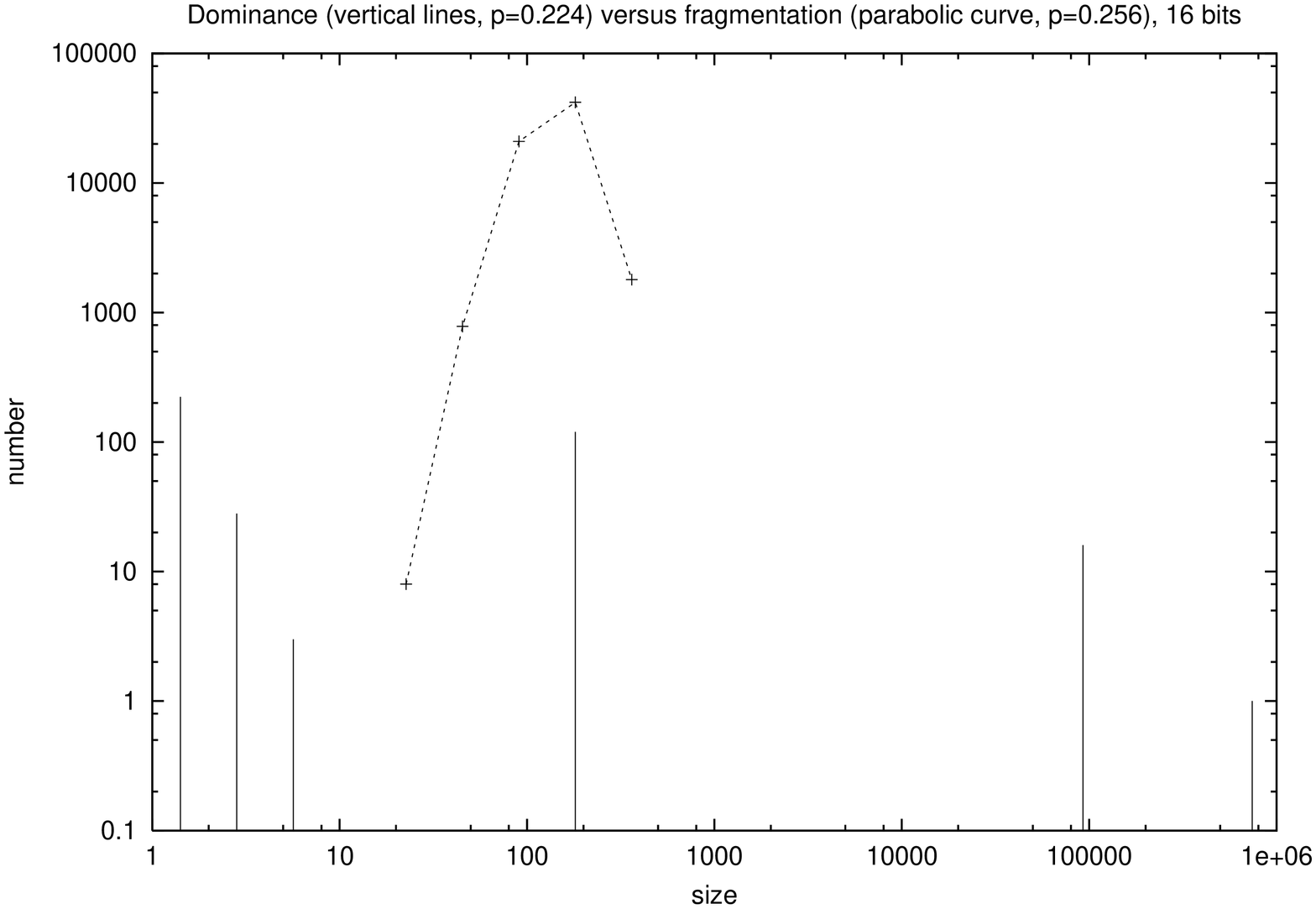}
\includegraphics[angle=-90,scale=0.40]{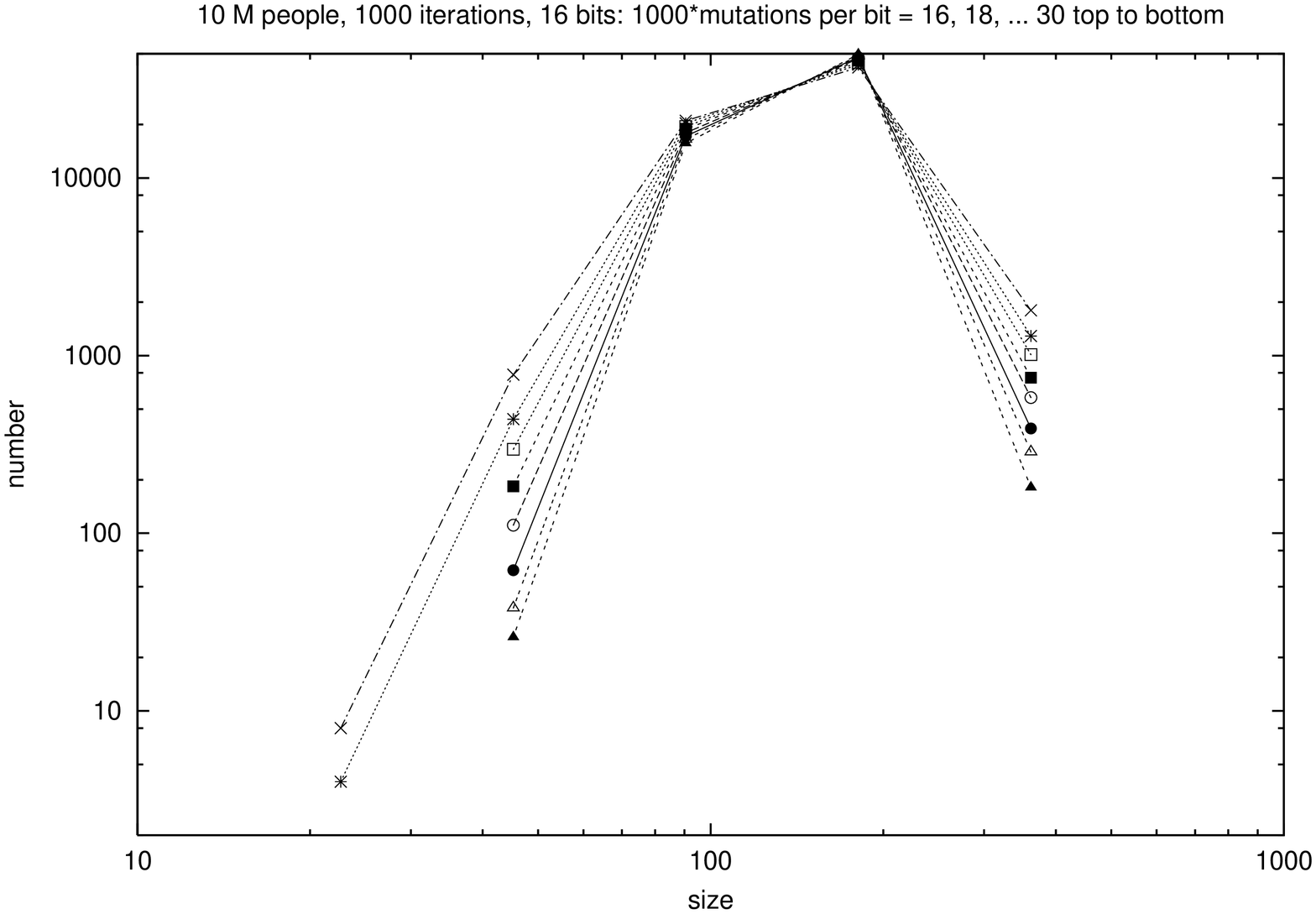}
\end{center}
\caption{ 
Histograms $P(S)$ of language sizes for 16 bits, one sample only of
$K/2 = 10$ million people, mutations per bit = 0.014 (bars) and 0.016 (parabola)
in part a and 0.016 to 0.030 in steps of 0.002 in part b from \cite{schulze}.
We see a wide distribution for dominance (vertical bars in part a) and a 
narrow one for fragmentation (part b). 
}
\end{figure}

\begin{figure}[hbt]
\begin{center}
\includegraphics[angle=-90,scale=0.35]{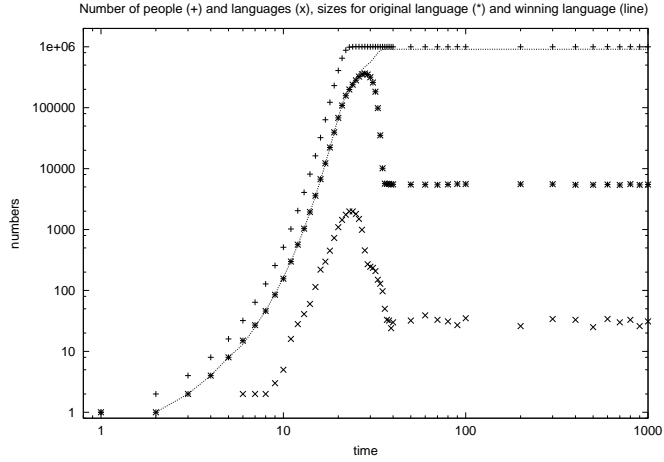}
\end{center}
\caption{Time development of the population (+) from one to one million people.
The number of languages spoken by at least 10 people is denoted by x, the 
number of speakers of the original language zero by stars, and the number of 
speakers of the dominating language by lines. The mutation rate per
person is 0.16 and the bit-strings have a length of $\ell = 16$. This dynamics
leads to dominance e.g. of the Roman alphabet and similar alphabets. See Nettle
\cite{comrie} for a similar plot of languages versus time.}
\end{figure}

\begin{figure}[hbt]
\begin{center}
\includegraphics[angle=-90,scale=0.35]{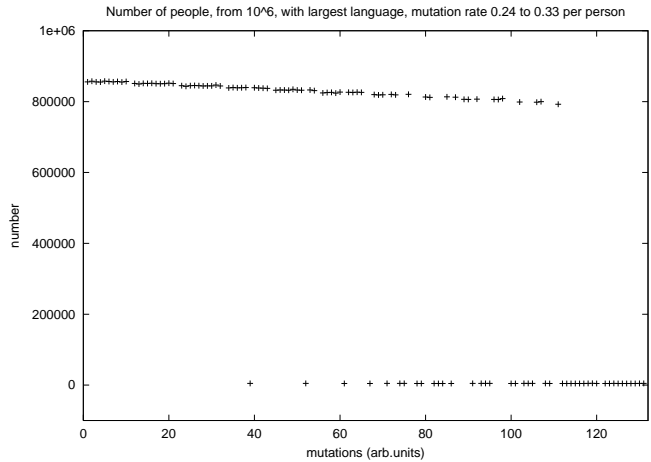}
\end{center}
\caption{Variation of size of the largest language with the mutation rate;
$\ell = 8$. The small sizes are slightly below 5000. We see a first-order
phase transition.
}
\end{figure}

\begin{figure}[hbt]
\begin{center}
\includegraphics[angle=-90,scale=0.35]{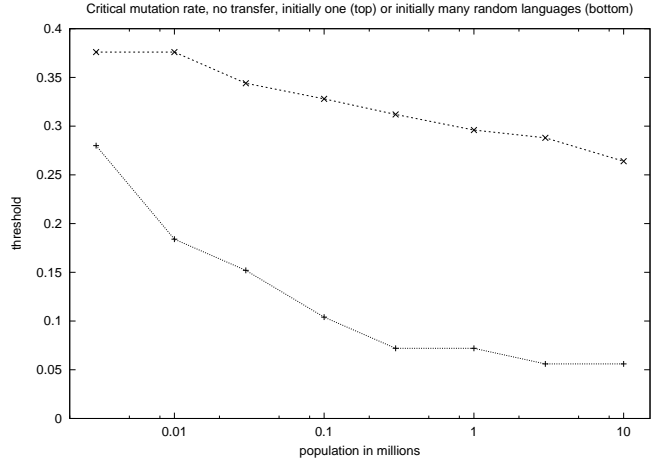}
\end{center}
\caption{Variation of the position of the phase transition with population
size for $\ell = 8$. We start
with the equilibrium population $K/2$ and either give all of them the same
language (upper curve) or each person a randomly selected language (lower 
curve). Below the curve, one language dominates; above it, many languages are
about equally important.  
}
\end{figure}

\begin{figure}[hbt]
\begin{center}
\includegraphics[angle=-90,scale=0.35]{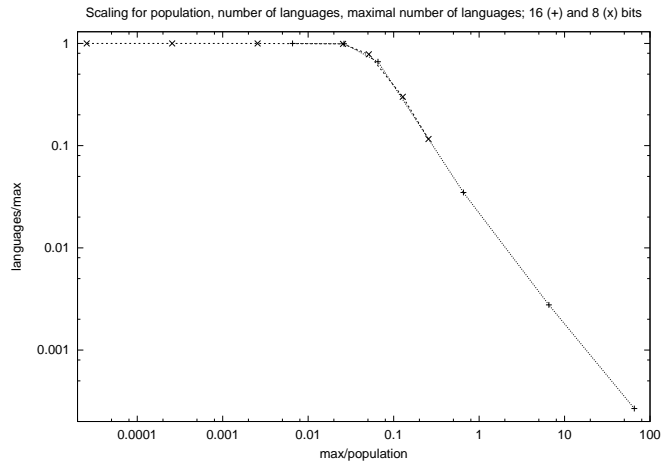}
\end{center}
\caption{ 
Scaling test: Symbols for $\ell = 8$ (x) and 16 (+) follow the same scaling
function $f$ if plotted as $L/M$ versus $M/N$. Also runs with 32 and 64
bits show the power-law decay of the right part. From Ref.\cite{schulze}.
(Languages spoken by less than 10 people are not counted here.) 
}
\end{figure}

\begin{figure}[hbt]
\begin{center}
\includegraphics[angle=-90,scale=0.30]{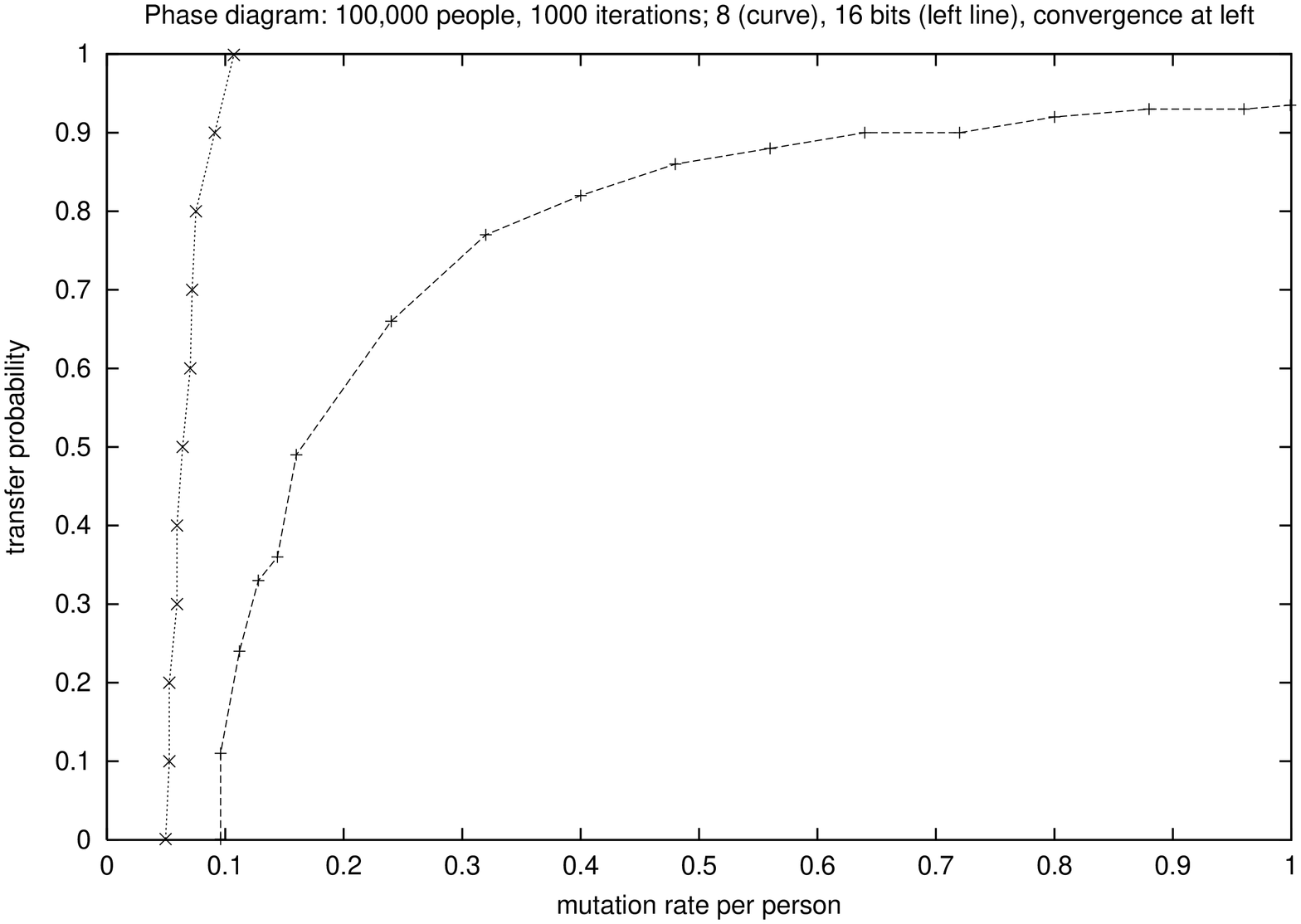}
\includegraphics[angle=-90,scale=0.30]{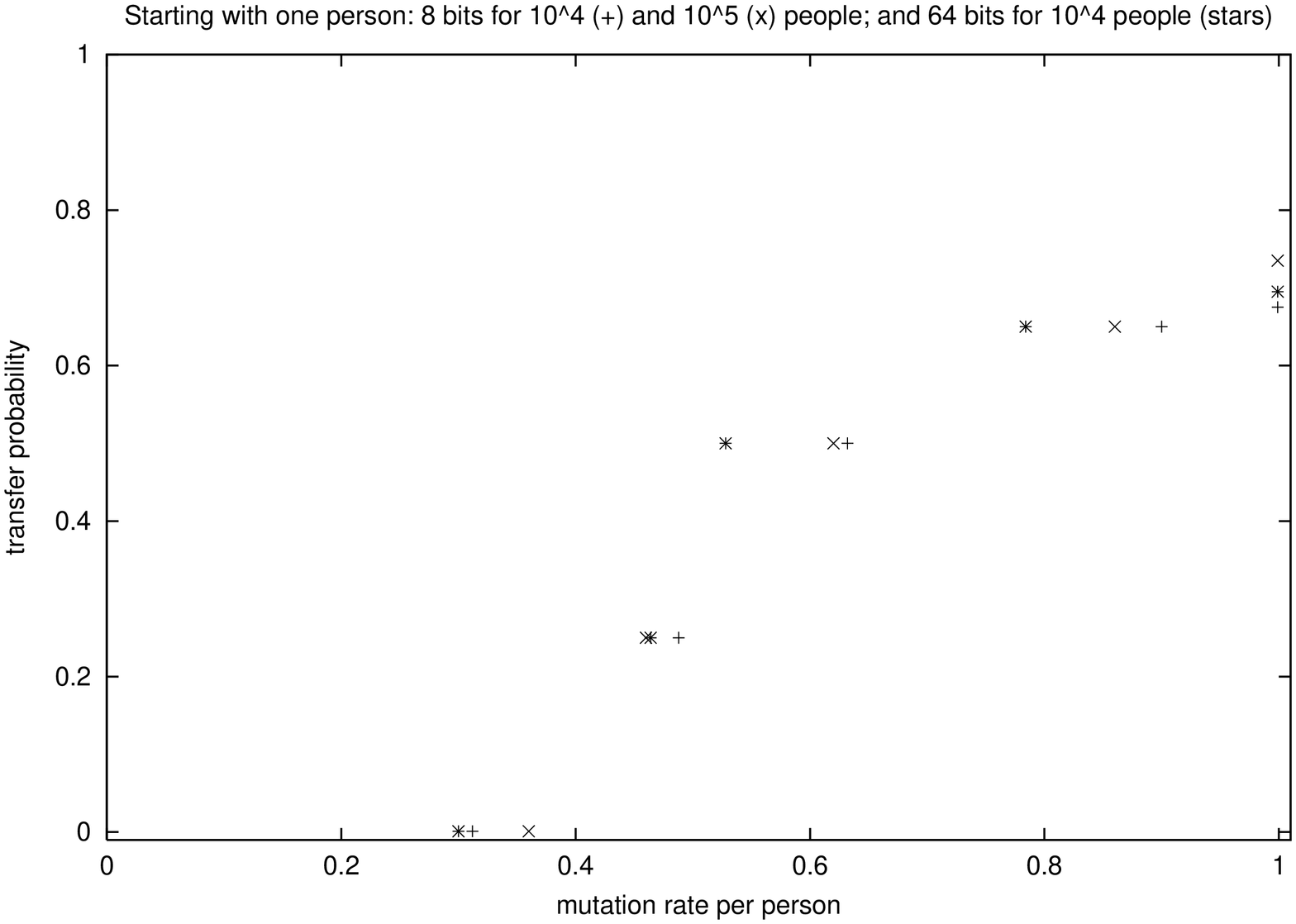}
\includegraphics[angle=-90,scale=0.30]{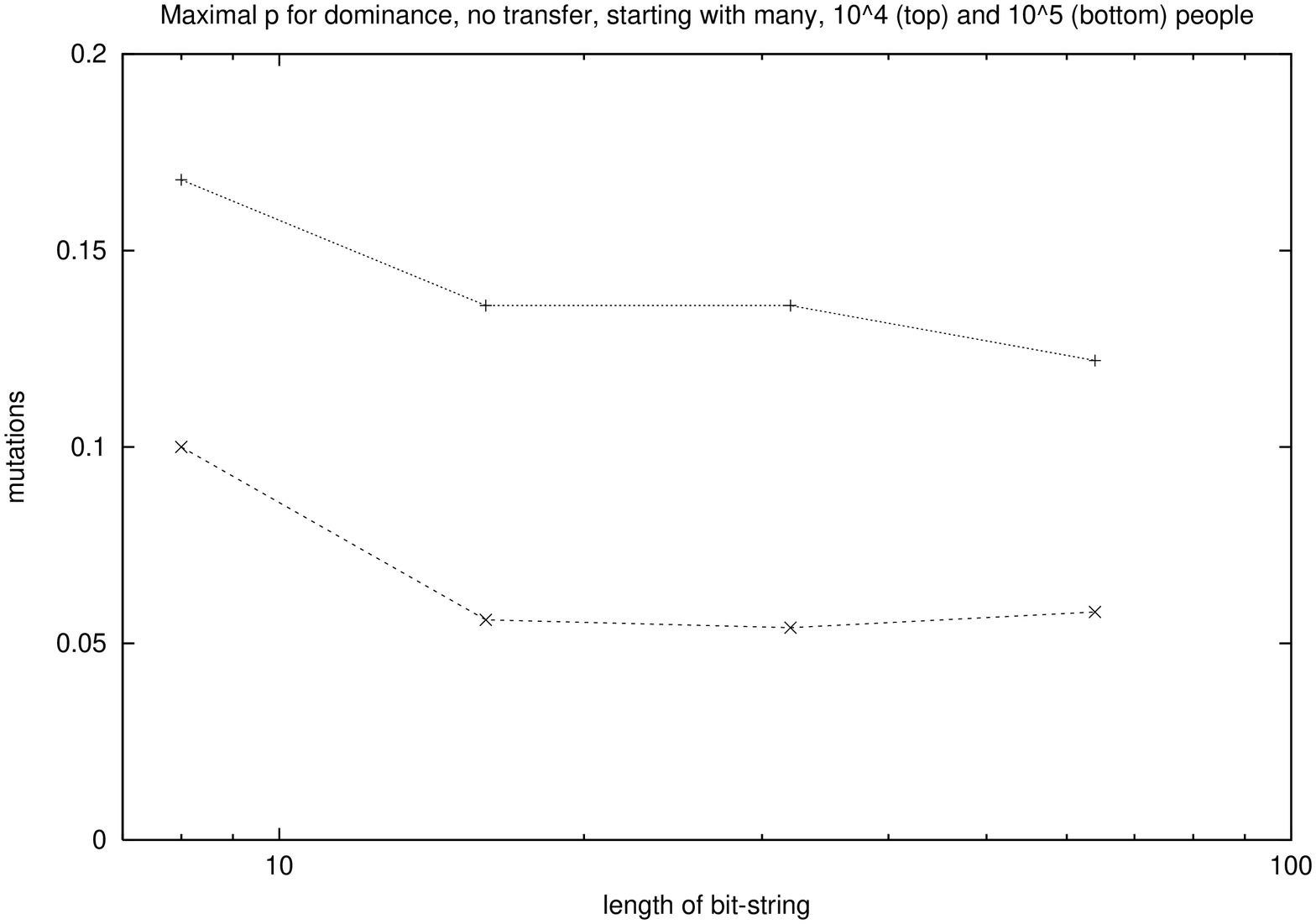}
\end{center}
\caption{Phase diagram in the $p-q-$plane. Part a starts with many languages
using $\ell=8$ and 16, while Part b starts with one person and uses 
$\ell$ up to 64. 
Fragmentation occurs in the lower right corner of parts a and b, dominance
in the upper left corner.
Part c shows that more than 16 bits
do not change here much (starting with many as in part a; $q = 0.001$).
}
\end{figure}

\begin{figure}[hbt]
\begin{center}
\includegraphics[angle=-90,scale=0.40]{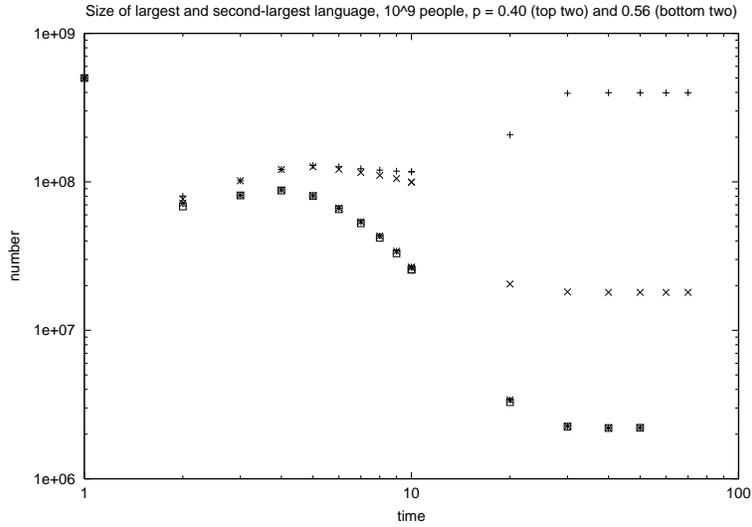}
\end{center}
\caption{Simulations with $10^9$ people for the merging model with two
languages initially at $q=0$, showing 
dominance for small and many languages for large $p$.
}
\end{figure}

\begin{figure}[hbt]
\begin{center}
\includegraphics[angle=-90,scale=0.35]{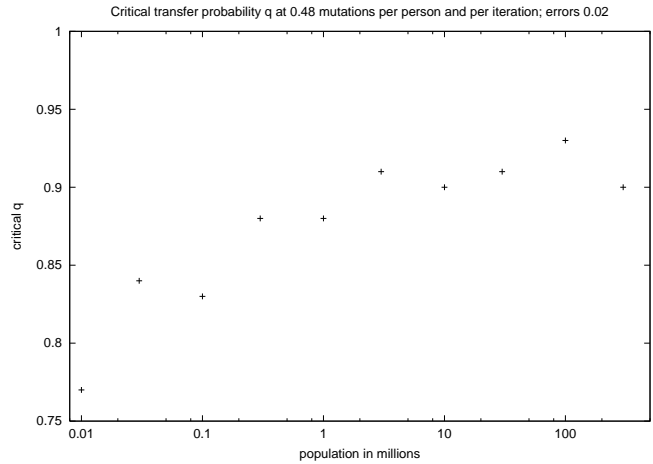}
\end{center}
\caption{ 
Strong size effects beyond the statistical error of about 0.02 in the
minimum transfer probability $q_c$ to achieve dominance of one language;
$\ell =8$. Perhaps for infinite population the critical $q$ becomes unity.
}
\end{figure}

\begin{figure}[hbt]
\begin{center}
\includegraphics[angle=-90,scale=0.4]{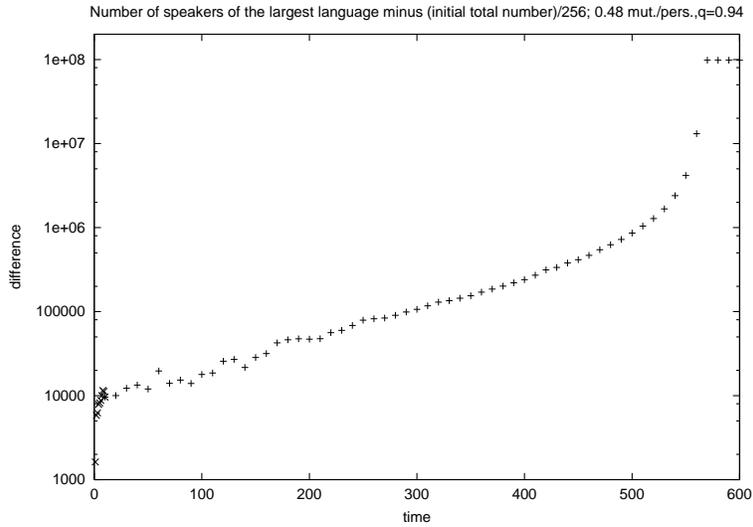}
\end{center}
\caption{ 
Roughly exponential increase of the advantage of the largest language;
the average size of all 256 languages ($\ell = 8$) is subtracted.
}
\end{figure}

\begin{figure}[hbt]
\begin{center}
\includegraphics[angle=-90,scale=0.4]{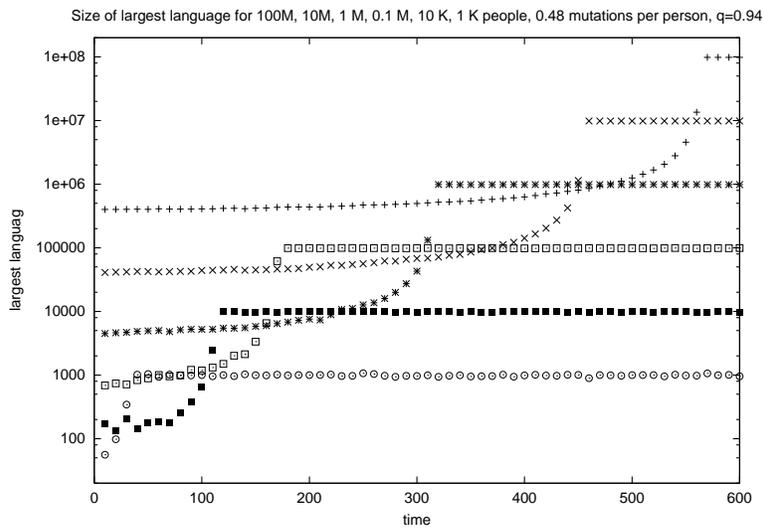}
\end{center}
\caption{ 
The larger the population is the longer we have to wait until one language
starts to dominate; $\ell = 8$.
}
\end{figure}

\begin{figure}[hbt]
\begin{center}
\includegraphics[angle=-90,scale=0.35]{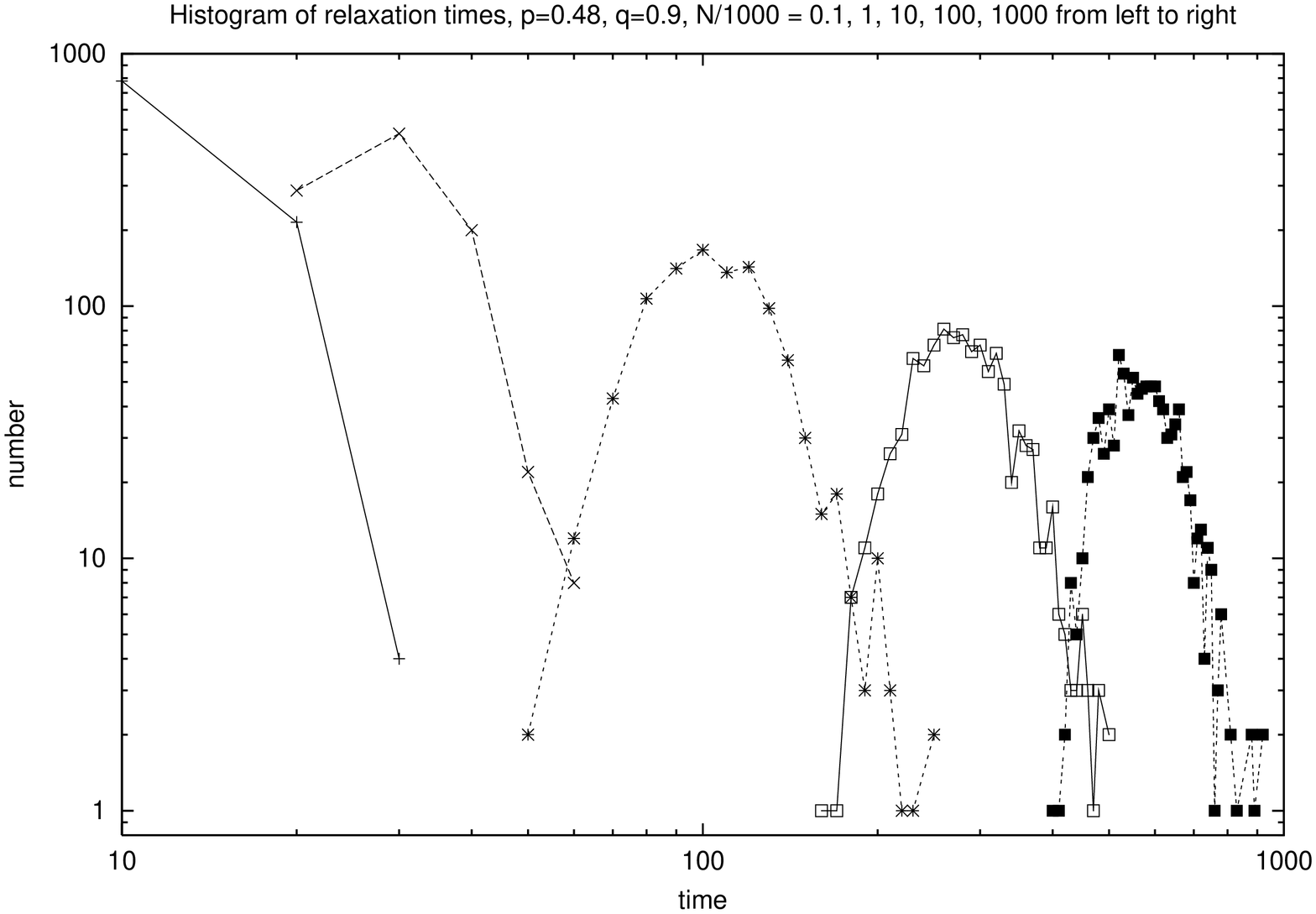}
\includegraphics[angle=-90,scale=0.35]{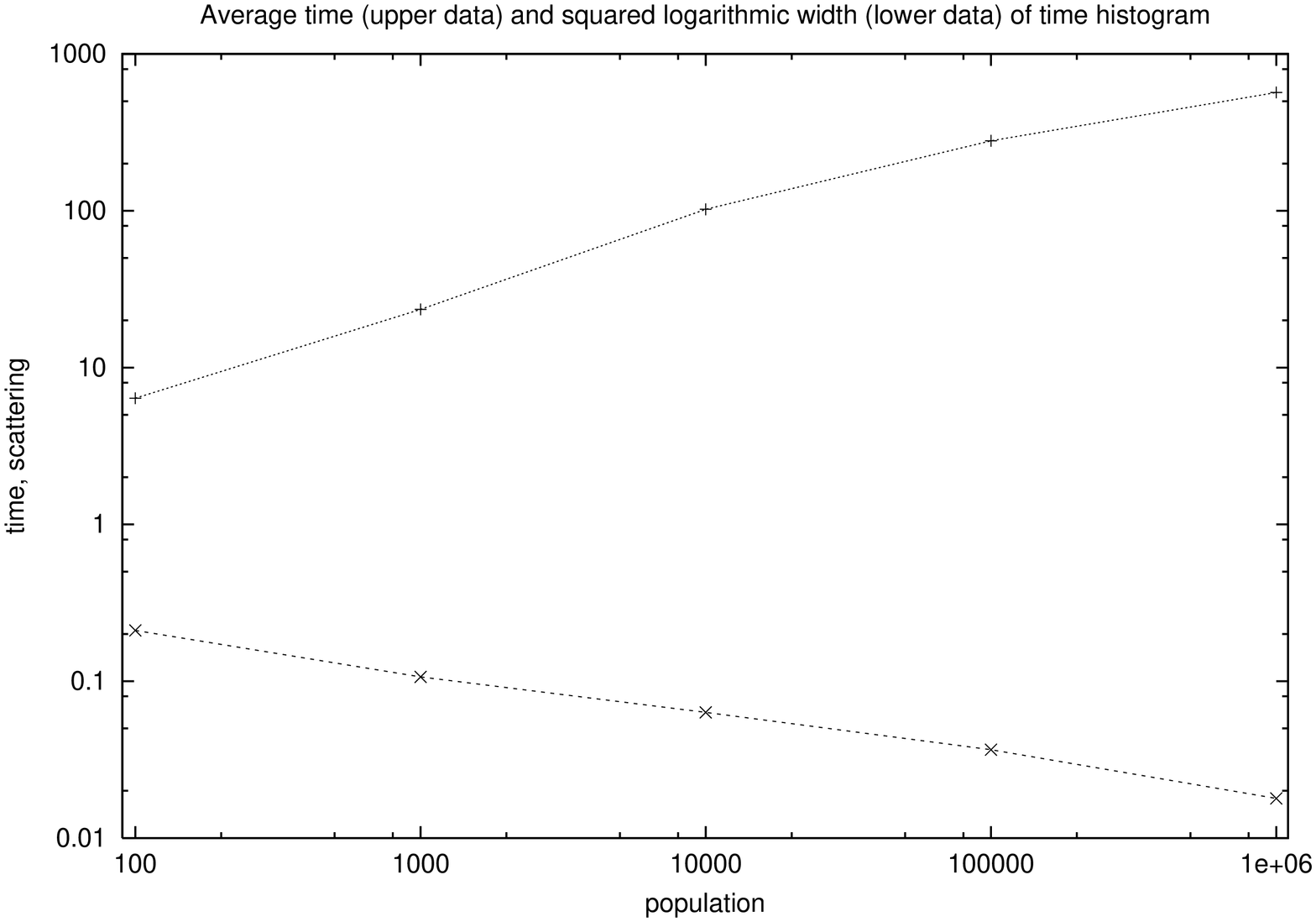}
\end{center}
\caption{ 
a) Roughly log-normal distribution of times after which one language is spoken
by at least half the people, for population increasing from left to right,
and $\ell = 8$. b) Resulting averages exp($<\ln t>$) and $<(\ln t)^2> - 
<\ln t>^2$. Thus for large populations, we get a narrow log-normal distribution
of the time to reach dominance, and the average of this time becomes large.
}
\end{figure}

\begin{figure}[hbt]
\begin{center}
\includegraphics[angle=-90,scale=0.40]{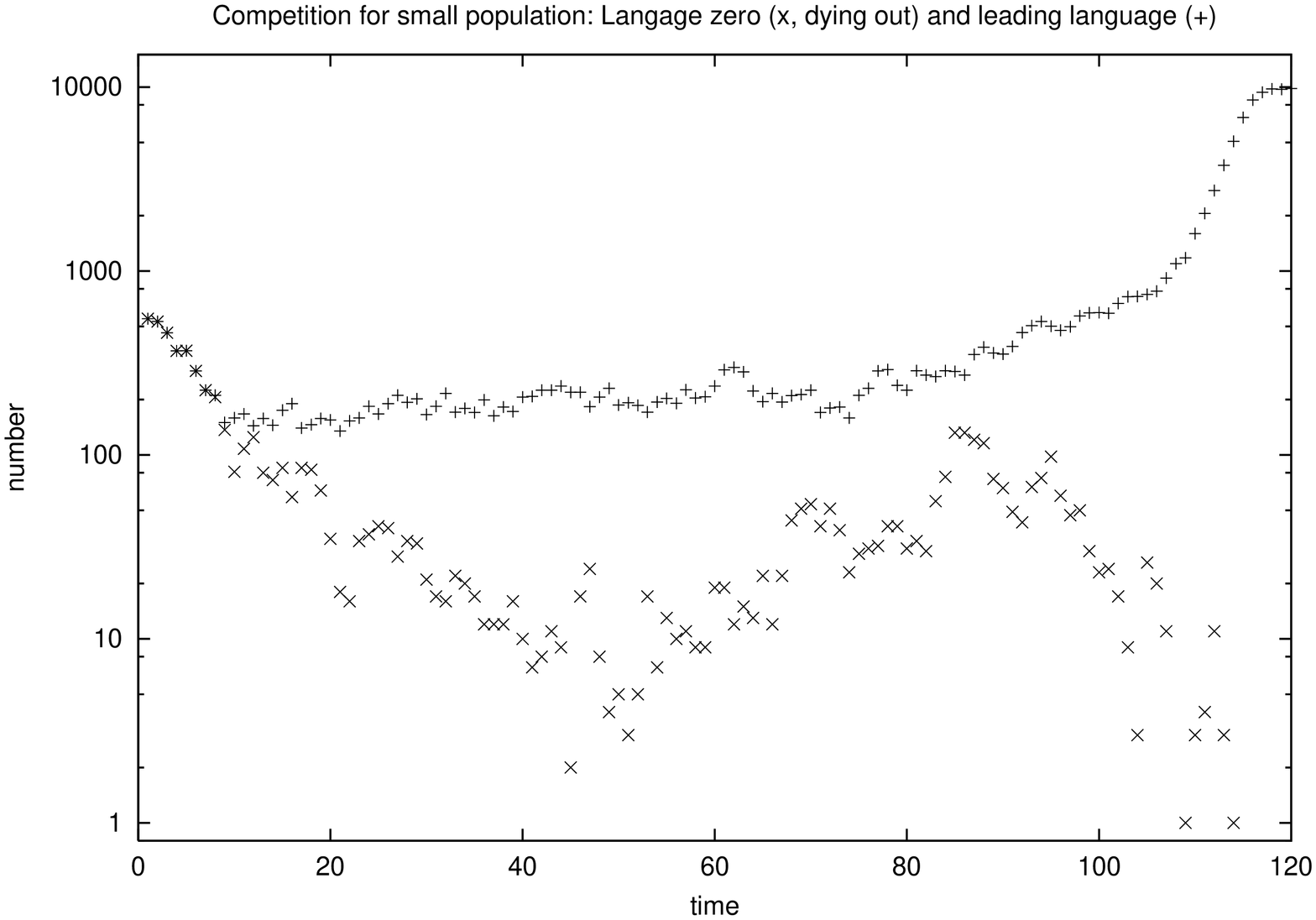}
\includegraphics[angle=-90,scale=0.40]{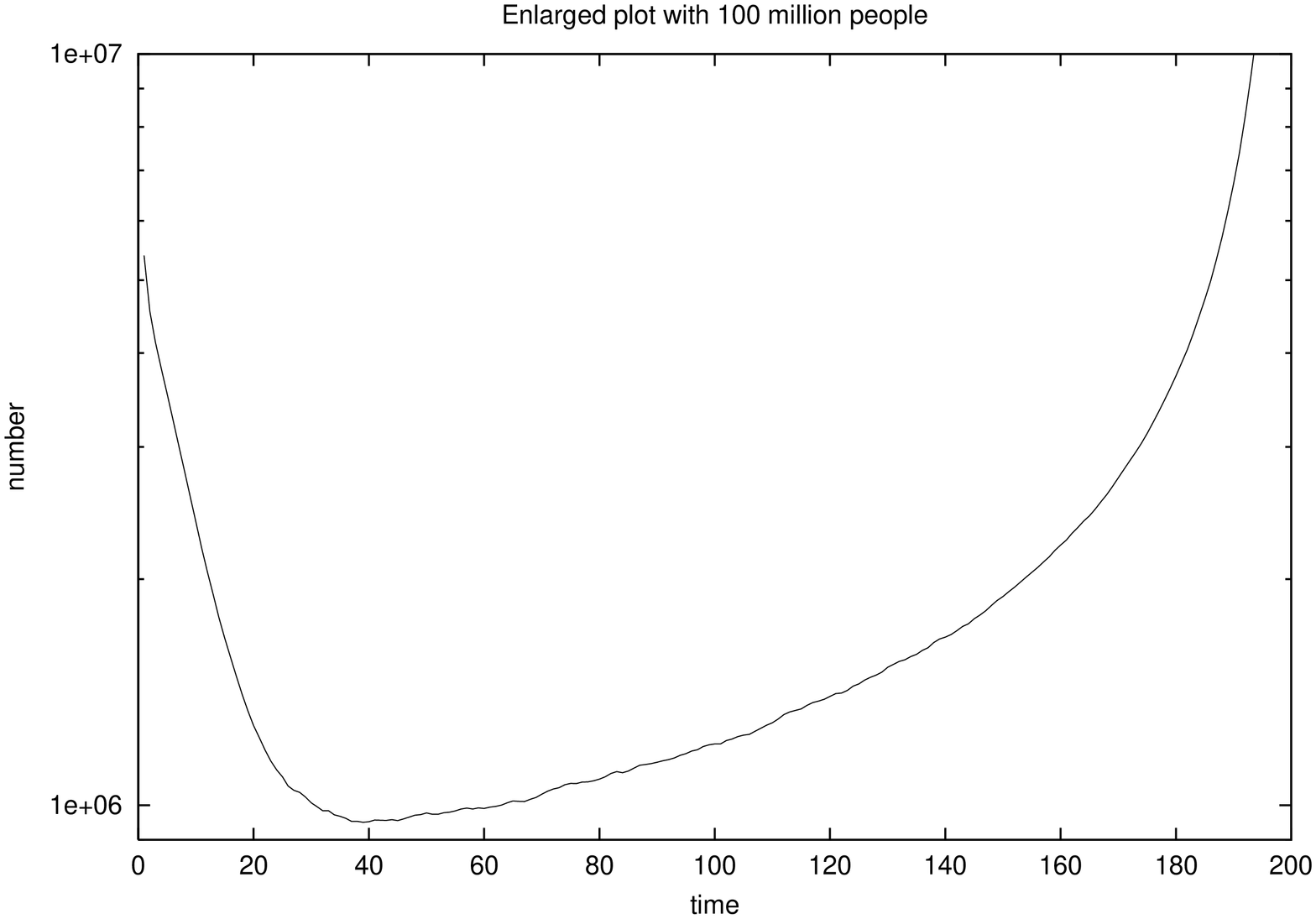}
\end{center}
\caption{ 
Does an initial advantage of 5 \% of the total population help the language
zero to win ? Not necessarily so for the small population of $10^4$ shown
in part a but clearly so for $10^8$ people in part b. $\ell = 8.$
}
\end{figure}

\begin{figure}[hbt]
\begin{center}
\includegraphics[angle=-90,scale=0.35]{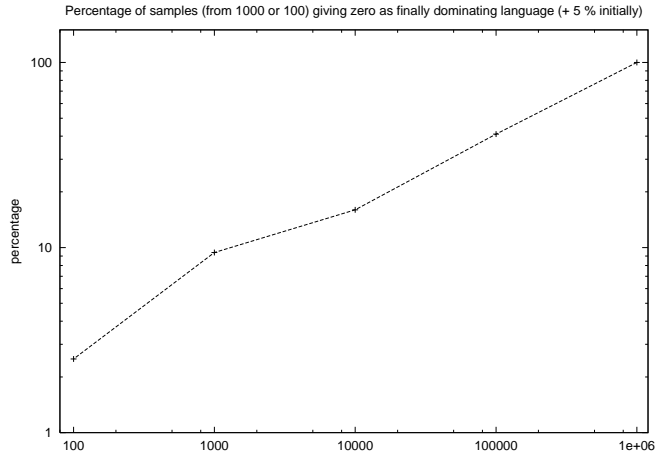}
\end{center}
\caption{ 
Percentage of samples where the initially favoured language wins; $\ell = 8$. 
}
\end{figure}

\begin{figure}[hbt]
\begin{center}
\includegraphics[angle=-90,scale=0.35]{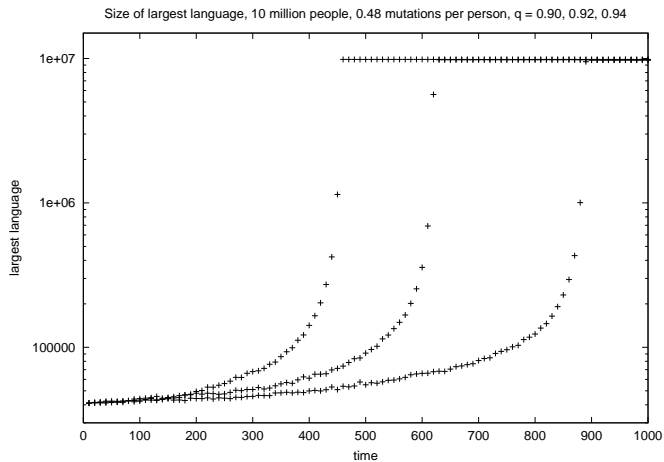}
\end{center}
\caption{ 
The closer we are to the phase transition the longer we have to wait for
dominance of one language to appear. The transfer probability $q$ is 0.90
for the rightmost curve, 0.92 for the centre curve, and 0.94 for the leftmost 
curve; $\ell = 8$.
}
\end{figure}

\begin{figure}[hbt]
\begin{center}
\includegraphics[angle=-90,scale=0.5]{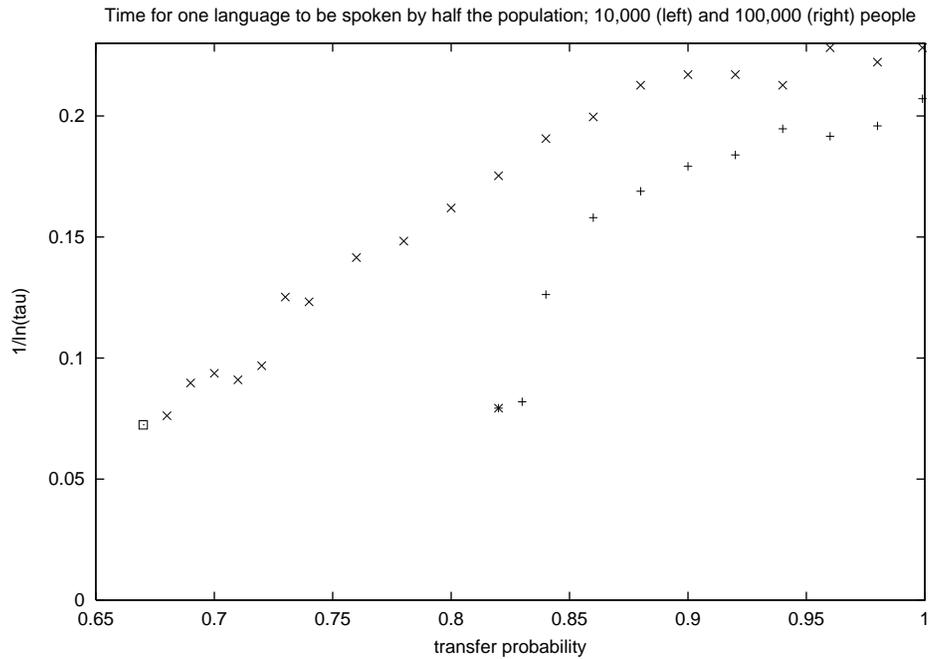}
\end{center}
\caption{ 
Indication that the time $\tau$ for dominance to appear diverges  at some
positive $q$ and not at $q=0$. We plot $1/\ln(\tau)$ versus $q$ and extrapolate
the data to hit the horizontal axis at a finite critical value  of $q$, which
strongly depends on the population size. (We
start with many randomly selected languages; $\ell = 8, \; p=0.48$. The 
star and square symbols are upper limits, i.e. up to that time no dominance was 
reached.)
}
\end{figure}

\clearpage
\begin{figure}[hbt]
\begin{center}
\includegraphics[angle=-90,scale=0.46]{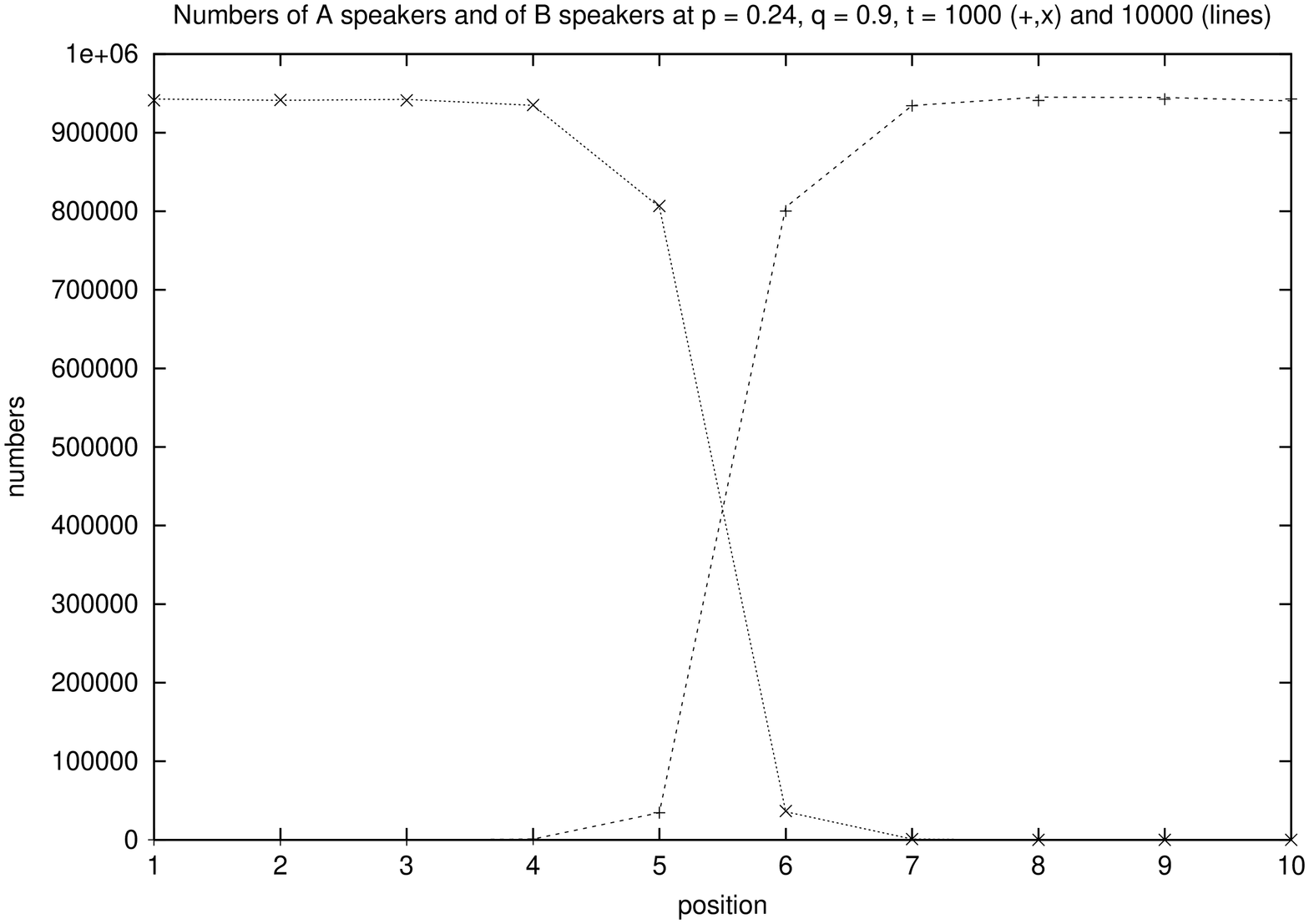}
\includegraphics[angle=-90,scale=0.46]{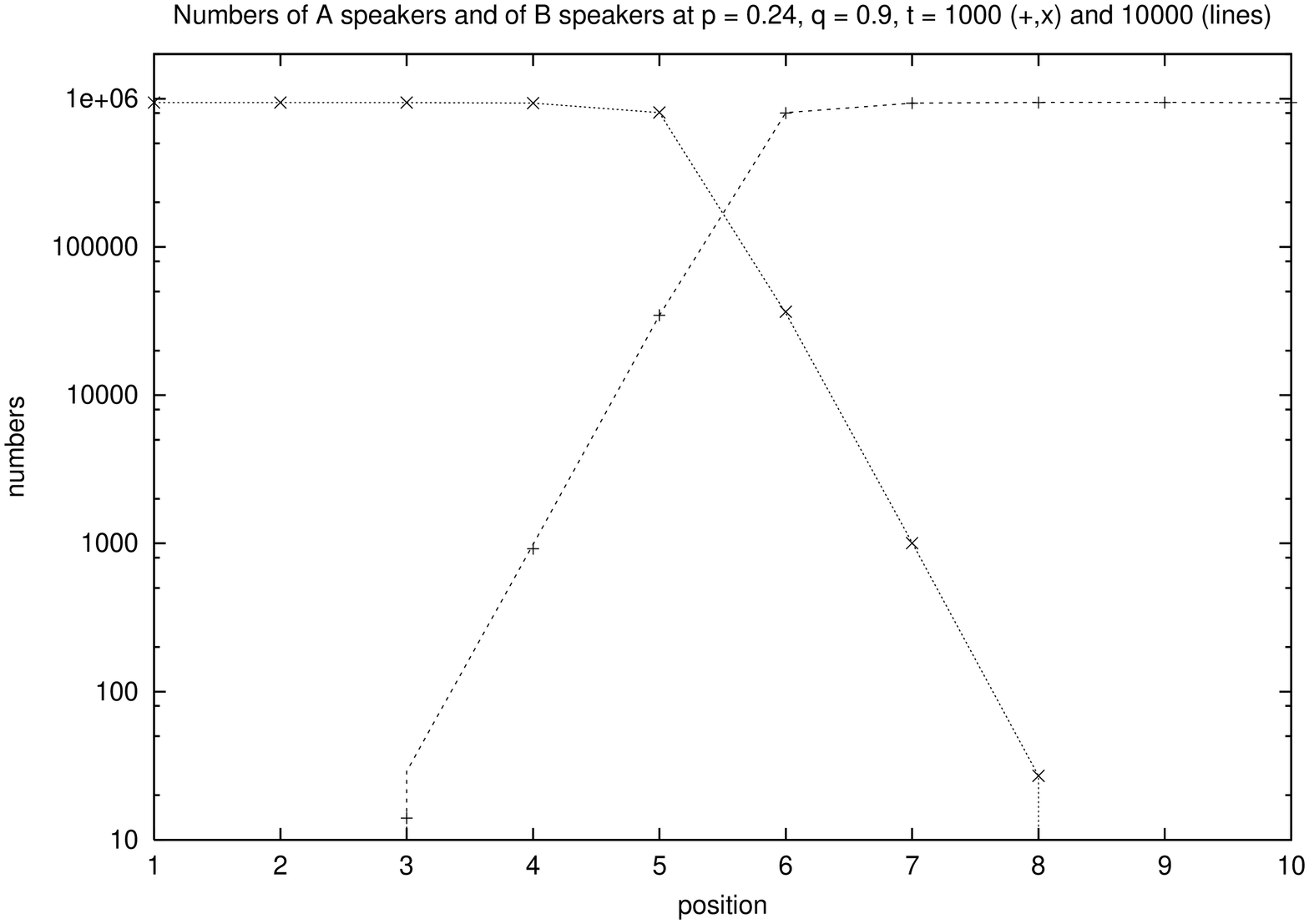}
\end{center}
\caption{Geographical language profiles near an interface.
Part a uses a linear population scale, part b a logarithmic one.
10 million people were distributed over a $10 \times 10$
square lattice. The interface width is about the same for $8 \times 8$,
$12 \times 12$ and $20 \times 20$.}   
\end{figure}

\begin{figure}[hbt]
\begin{center}
\includegraphics[angle=-90,scale=0.50]{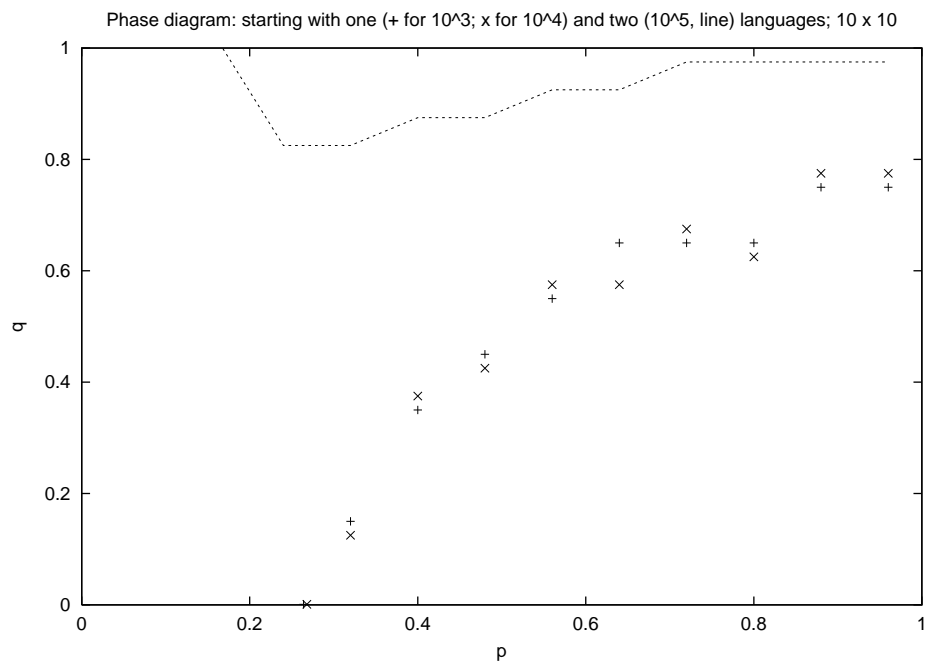}
\end{center}
\caption{Stability limit for initial configuration of one (bottom) or two (top)
languages in $10 \times 10$ lattice with $10^3 \dots 10^5$ individuals per
site.}   
\end{figure}

\end{document}